\documentclass[11pt]{article}
\usepackage{jheppub}
\usepackage{color}
\usepackage{array}
\usepackage{amsmath}
\usepackage{amssymb}
\usepackage{graphicx}

\makeatletter

\providecommand{\tabularnewline}{\\}

\numberwithin{equation}{section}

\@ifundefined{date}{}{\date{}}

\makeatother

\begin{document}
\title{Non-Perturbative Heterotic Duals of M-Theory on $G_{2}$ Orbifolds}
\author[a,b]{Bobby Samir Acharya,}
\author[c]{Alex Kinsella,}
\author[c,d]{and David R. Morrison}
\affiliation[a]{Department of Physics, Kings College London, London, WC2R 2LS, UK}
\affiliation[b]{Abdus Salam International Centre for Theoretical Physics, Strada Costiera 11, 34151, Trieste, Italy}
\affiliation[c]{Department of Physics, Broida Hall, University of California, Santa Barbara, CA 93106}
\affiliation[d]{Department of Mathematics, South Hall, University of California, Santa Barbara, CA 93106}

\begin{flushright}
UCSB-Math-2021-02 \\
KCL-PH-TH/2021-38
\end{flushright}

\abstract{By fibering the duality between the $E_{8}\times E_{8}$ heterotic
string on $T^{3}$ and M-theory on K3, we study heterotic duals of
M-theory compactified on $G_{2}$ orbifolds of the form $T^{7}/\mathbb{Z}_{2}^{3}$.
While the heterotic compactification space is straightforward, the description of the
gauge bundle is subtle, involving the physics of point-like instantons
on orbifold singularities. By comparing the gauge groups of the dual
theories, we deduce behavior of a ``half-$G_{2}$'' limit, which is
the M-theory analog of the stable degeneration limit of F-theory.
The heterotic backgrounds exhibit point-like instantons that are localized
on \emph{pairs} of orbifold loci, similar to the ``gauge-locking''
phenomenon seen in Ho\v{r}ava--Witten compactifications. In this way, the
geometry of the $G_{2}$ orbifold is translated to bundle data in
the heterotic background. While the instanton configuration looks
surprising from the perspective of the $E_{8}\times E_{8}$ heterotic
string, it may be understood as T-dual $\text{Spin}(32)/\mathbb{Z}_{2}$
instantons along with winding shifts originating in a dual Type I
compactification.}

\maketitle

\section{Introduction and Summary}

One of the well-known dualities in string theory relates M-theory
compactified on a K3 surface to the $E_{8}\times E_{8}$ heterotic
string compactified on a three-torus \cite{Hull:1994ys,Witten:1995ex}. It was proposed long ago that
this 7D M/heterotic duality could be applied fiberwise over an $S^{3}$
base to obtain a 4D duality as well \cite{Acharya:1996ci,Acharya:2001gy,gukov2003duality}.
In this case, M-theory is compactified on a $G_{2}$ manifold equipped
with a coassociative K3 fibration, while the $E_{8}\times E_{8}$
heterotic string is compactified on a Calabi--Yau threefold equipped
with a supersymmetric three-torus fibration (also known as an SYZ
fibration \cite{Strominger:1996it}).

One way to exhibit the 7D M/heterotic duality is to take the large heterotic
volume limit, which corresponds to the ``half-K3'' limit on the
M-theory side \cite{morrison2002half}. There is a limiting family
of K3 metrics in which a long throat of the form $T^{3}\times I$
develops, where $I$ is an interval, and the complicated geometry
is confined to the two ends. Each complicated end is known as a half-K3
surface and carries a metric known as an ALH instanton \cite{Biquard:2010ig}. These half-K3
surfaces each determine an $E_{8}$ bundle on $T^{3}$, together giving
a heterotic string gauge background \cite{morrison:limits}.

One can then attempt to find a similar fiberwise picture for a $G_{2}$
space $X$ with a coassociative K3 fibration. Under favorable conditions,
there will be a family of metrics in which a long throat of the form
$Y\times I$ develops, where $Y$ is the SYZ-fibered Calabi--Yau
threefold appearing as the heterotic dual. We call this the ``half-$G_{2}$''
limit, and in this paper we will discuss aspects of M/heterotic duality
in this limit that go beyond the perturbative picture of the half-K3
limit. Our\textcolor{red}{{} }goal is to work towards a dictionary between $G_{2}$ spaces and the heterotic gauge bundle. We approach this
task by trying to answer this question in the simple case of a Joyce orbifold: how is
the geometry of the ambient $G_{2}$ space reflected by the heterotic
bundle, which lives only on a suborbifold? For the simple examples
studied in this paper, the topological data on the $G_{2}$ side is
captured by the configuration of the orbifold singular loci and their
intersections with codimension-$1$ suborbifolds. This data is spread
throughout the throat interval in the half-$G_{2}$ limit, as opposed
to the situation of the half-K3 limit, where the singularities are
confined to the ends of the interval. On the heterotic side, this
data is represented by point-like instantons on orbifold singularities.
We find point-like instanton configurations that look somewhat exotic
from the $E_{8}\times E_{8}$ perspective, but can be understood as
T-dual $\text{Spin}(32)/\mathbb{Z}_{2}$ point-like instantons on
an orbifold with a winding shift. 

In general, M/heterotic duality shares many properties with heterotic/F-theory
duality, and in some cases the two are directly related via a duality
chain. This duality was used in \cite{Braun:2017uku} to study M-theory
on twisted-connected sum $G_{2}$ spaces that support fibrations by
K3 surfaces that are themselves elliptically fibered. Beyond the twisted-connected
sum examples, a generic compactification of M-theory on a K3-fibered
$G_{2}$ space is not expected to have an F-theory dual, and must
be studied in terms of differential geometry instead of complex geometry.
In this paper we
explore M/heterotic duality without the tools of elliptic fibrations
on the M-theory side. One useful perspective in this case is duality
with the Type I string, where tadpole cancellation conditions give
additional computational tools. 

It has long been recognized that M-theory needs to be compactified
on spaces with singularities in order to produce interesting gauge
groups and matter content in the effective theory \cite{Acharya:1998pm,Acharya:2001gy}.
Joyce's work \cite{10.4310/jdg/1214458109,10.4310/jdg/1214458110}
is celebrated for demonstrating the existence of nonsingular compact manifolds
with holonomy $G_{2}$, but ironically, the singular $T^{7}/\Gamma$
orbifolds from which Joyce started are more relevant to the physics
than their nonsingular cousins. Those orbifolds have flat metrics
and a natural $G_{2}$ structure encoded in an invariant three-form,
which is the limit of the smooth $G_{2}$ structures when the resolved
singularities are blown back down. In this paper we will study those
orbifolds themselves. The resulting effective theories preserve $N=1$
supersymmetry and have ADE gauge groups, but the lack of codimension
$7$ singularities implies that there is no chiral matter, so that these particular
Joyce orbifolds cannot produce phenomenologically realistic effective
theories in this limit. However, these orbifolds produce a simple laboratory within
which to deduce properties of duality that are expected to persist
for more realistic examples. 

In many of Joyce's orbifolds, there is a fibration by flat Kummer
surfaces of the form $T^{4}/\mathbb{Z}_{2}$. It is precisely in such an
orbifold limit that Ricci-flat metrics on K3 surfaces are easy to
construct, because in that limit those metrics are flat. The corresponding
fibration is by coassociative cycles of $T^{7}/\Gamma$, with $\Gamma$
a finite group, and again the coassociative condition is trivial to
check because we are working with flat metrics\footnote{It is an open question whether in Joyce's resolution of singularities
there are smooth K3 surfaces which resolve the singularities of the
Kummer surfaces in such a way as to form a coassociative fibration.}. The geometry of Kummer fibrations of $G_{2}$ orbifolds was analyzed
in detail by Liu \cite{Liu:1998tha}, whose work forms part of the foundation
upon which we develop heterotic duals. 

To find the half-$G_{2}$ limit, we identify a particular $S^{1}\subset T^{7}$
on which $\Gamma$ acts as a reflection, so that there is a fibration
$T^{6}/H \to T^{7}/\Gamma \to S^{1}/\mathbb{Z}_{2}$ with $H$ a subgroup of $\Gamma$ and the ends of the interval
$S^{1}/\mathbb{Z}_{2}$ the location of the complicated geometry.
In all of the examples we consider, the Calabi--Yau threefold $Y$
is also an orbifold $T^{6}/H$, and in our $N=1$ supersymmetric cases, it is an
orbifold of a special type known as a Borcea--Voisin orbifold\footnote{One of the advantages of this observation is that Gross and Wilson
analyzed SYZ fibrations on Borcea--Voisin orbifolds and on their
resolutions \cite{gross:1997}.} \cite{Borcea:1997tq,voisin1993miroirs}. In fact, our $N=1$ examples
all live on the same Borcea--Voisin orbifold, which is the blow-down
limit of the Schoen manifold, in agreement with the results of \cite{Braun:2017uku}. 

Identification of the heterotic dual requires specifying a background
gauge bundle with connection on the heterotic Calabi--Yau $Y$, which is $T^{6}/H$ or its resolution. Ideally,
we would have an algorithmic procedure to determine this bundle from
the M-theory data, in analogy to the case of heterotic/F-theory duality
\cite{Friedman:1997yq}, but this is made difficult by the fact that
the $T^{3}$ fibers of $Y$ are not complex submanifolds, so we must
instead identify
the dual bundle by indirect means. One useful tool is the
matching of massless spectra on the two sides. In particular, we may
split the heterotic spectrum into a perturbative part and a non-perturbative
part, where the former may be seen from a CFT analysis, while the
latter comprises the effects that are non-perturbative in the (heterotic) string
coupling. These two parts of the dual heterotic spectrum are distinguished
on the M-theory side by whether individual components of the singular locus 
of the $G_{2}$
orbifold are transverse to the generic fiber of the K3 fibration or not,
in the spirit of \cite{Bershadsky:1996nh}. The split refines our
analysis of the dual pair, as we must ensure that the heterotic particles
have the correct perturbative/non-perturbative origin. 

The perturbative spectrum may be obtained by breaking of primordial gauge symmetry by the monodromy of instanton connections sitting on the orbifold singularities. We expect the non-perturbative part of the heterotic spectrum to come
from these instantons in the singular point-like limit. Such gauge configurations
are consistent with heterotic anomaly cancellation conditions and
are the best-understood sources of non-perturbative gauge symmetry
in heterotic $E_{8}\times E_{8}$ compactifications. The massless
particle contributions of point-like instantons are partially understood
in simple examples, but distinguishing between different cases can
be subtle \cite{Aspinwall:1998he}, and there is no complete classification.
Some of the point-like instantons that we identify in dual heterotic
backgrounds are supported on pairs of orbifold loci and do not look familiar from
previous studies of point-like instantons on orbifold singularities.
This may be an analog of the gauge locking phenomenon seen in Ho\v{r}ava--Witten
compactifications \cite{Kaplunovsky:1999ia,Gorbatov:2001pw,Marquart:2002bz}
or a freezing of heterotic moduli by a gauge bundle configuration
\cite{Anderson:2011ty}. In the non-singular limit, candidate local descriptions for this type of instanton may be given by $\mathbb{Z}_2$-quotients of instantons on $\mathbb{R}^4$ or a caloron on $\mathbb{R}^3 \times S^1$ \cite{Lee:1998vu,Kraan:1998pm}. The behavior of the point-like instantons is more
clear from a T-dual $\text{Spin}(32)/\mathbb{Z}_{2}$ 
perspective \cite{Berkooz:1996iz},
where the background is acted upon by a winding shift.  

This paper is organized as follows. Section 2 gives an overview of
the fundamental M/heterotic duality in 7D and its fibration over a
3D base. Section 3 discusses M-theory on $G_{2}$ orbifolds and analyzes
three examples of K3-fibered $G_{2}$ orbifolds that will form the
heart of the paper. In section 4, we examine the dual heterotic geometry,
a Borcea--Voisin orbifold, that is dictated by the duality in the half-$G_{2}$
limit. In section 5, we survey non-perturbative aspects of the heterotic
gauge bundle, and in particular point-like instantons on orbifold
singularities. This prepares us to analyze the gauge bundles of our
dual heterotic examples in section 6.\textcolor{red}{{} }In section
7, we investigate the nature of the heterotic gauge bundle via an
alternative duality chain relating our M-theory setup to Type I compactifications
on orbifolds with winding shifts. Finally, in section 8, we interpret
our results in terms of Ho\v{r}ava--Witten duals, gauge locking, and frozen
moduli and discuss future directions.

\section{Heterotic/M-Theory Duality}

\subsection{Duality in 7D}

To obtain dual low energy effective theories in 4D, we will make use
of the duality between the 7D theories arising from the $E_{8}\times E_{8}$ heterotic string on $T^{3}$ and M-theory on the compact 4-manifold known as a K3 surface 
\cite{Witten:1995ex}. Evidence for this duality comes in part from
the fact that these two compactifications share the same moduli 
space:\footnote{There  are some subtleties concerning the discrete group 
action which we suppress here.}
\[
{\cal M}_{7D}=\left[\text{SO}(3,19;\mathbb{Z})\backslash\text{SO}(3,19;\mathbb{R})/\text{SO}(3;\mathbb{R})\times\text{SO}(19;\mathbb{R})\right]\times\mathbb{R}^{+} \ .
\]
On the M-theory side, the first factor is interpreted as the moduli
space of volume-1 Einstein metrics on K3, while the $\mathbb{R}^{+}$
factor is the volume. On the heterotic side, the first
factor is instead interpreted as the Narain moduli space of heterotic
compactifications on $T^{3}$, while the $\mathbb{R}^{+}$ is the
string coupling. By comparing the effective actions on each side of
the duality, one finds the relation between the $\mathbb{R}^{+}$
factors 
\[
e^{3\gamma}=\lambda \ ,
\]
where $e^{3\gamma}$ is the volume of the K3 surface and $\lambda$
is the heterotic string coupling.

There are special points in the moduli space where non-abelian gauge
symmetry appears in the 7D theory. From the heterotic side, these
points are those at which the holonomy of the flat $E_{8}\times E_{8}$
connection over the $T^{3}$ is non-generic. The unbroken gauge symmetry
in the effective theory is given by the centralizer of the reduced
structure group of the gauge bundle with connection. In the case of
a flat connection, this is the centralizer of the holonomy group,
which is generated by three commuting elements of $E_{8}\times E_{8}$\footnote{In this paper, we only consider the identity-connected component of the space of flat connections. See \cite{deBoer:2001wca} for discussion of the other components.}.
For a generic choice of these three elements, the gauge symmetry is
reduced to the maximal torus $\text{U}(1)^{16}$, but non-generic
holonomies give instead ADE gauge groups. 

From the view of M-theory, the special points in the moduli space
are orbifold limits of K3 that contain ADE singularities \cite{Witten:1995ex}.
That these singularities give rise to effective non-abelian gauge
symmetry can be seen by blowing up an $A_{1}$ singularity to give
an exceptional $\mathbb{P}^{1}$: this cycle is dual to a harmonic
2-form, which gives an effective $\text{U}(1)$ gauge field upon Kaluza-Klein
reduction of the C-field. Wrapping two M2-branes of opposite orientation
on the cycle give effective vector particles charged under the $\text{U}(1)$.
As the $\mathbb{P}^{1}$ shrinks to zero volume, the charged particles
become massless and complete the $\mathfrak{su}(2)$ Lie algebra.
A similar argument extends to general ADE singularities. 

\begin{figure}
\begin{centering}
\includegraphics[scale=0.25]{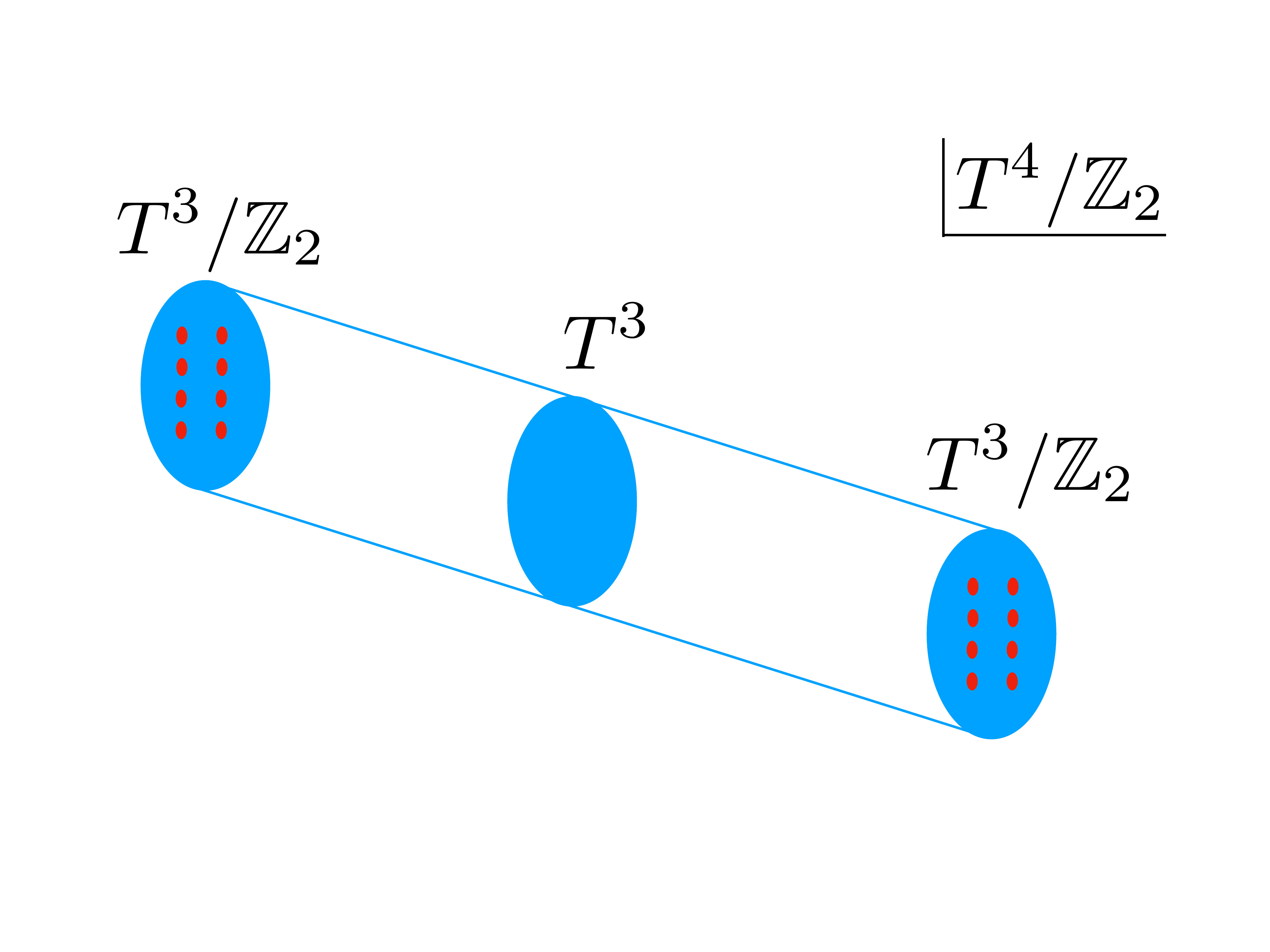}
\par\end{centering}
\caption{The half-K3 limit of $T^{4}/\mathbb{Z}_{2}$. The space degenerates
into a long throat with cross section $T^{3}$, while the 16 orbifold
points, which correspond to the complicated geometry of the resolved
space, recede to either end of the throat. If we put M-theory on this
space, then the dual heterotic theory lives on the central $T^{3}$
and has gauge bundle determined by the distant singularities. }
\end{figure}

\subsection{The Half-K3 and Weak Coupling Limits}

The heterotic string on $T^{3}$ has two primary dimensionless parameters:
the dimensionless compactification volume $\frac{\text{vol}T^{3}}{\alpha'^{3/2}}$
and the string coupling $\lambda$. Where possible, we will work in
the corner of the 7D parameter space where the compactification volume
is large and the string coupling is small. The large volume limit
is essential to current investigations into M/heterotic duality because
it is where we can differentiate the moduli corresponding to the heterotic
geometry and the gauge bundle, so that we may apply a geometric version
of the duality \cite{morrison2002half,Braun:2017uku}. The weak coupling
limit allows us to understand the heterotic physics via perturbation
theory combined with instanton effects. 

Both of these limits have a geometric realization on the M-theory
side. Large heterotic volume corresponds to what is called the ``half-K3
limit'' (see Figure 1): the K3 grows a long throat where the geometry
is slowly varying and approximately $T^{3}\times(-r,r)$ for some
$r\in\mathbb{R}$, so that all of the complicated geometry recedes
to $\pm r$ \cite{morrison2002half}. In this limit, the 7D duality
is realized by splitting the K3 surface in half, cutting transverse
to the throat. This gives us two 4-manifolds with $T^{3}$ boundary
-- these are ``half-K3 surfaces''. Such a surface may be realized
as a rational elliptic surface with a generic divisor (an elliptic
curve) removed. The dual heterotic theory is compactified on the $T^{3}$
boundary shared by the half-K3 surfaces. The geometry of these surfaces
contains the data for the $E_{8}\times E_{8}$ heterotic gauge bundle
on $T^{3}$. Specifically, the moduli of a half-K3 together with an
embedded $T^{3}$ is the same as the moduli of an $E_{8}$ bundle
on $T^{3}$. This half-K3 limit is analogous to the stable degeneration
limit of 8D F-theory/heterotic duality, where large volume of the
heterotic $T^{2}$ is dual to a limit in which the F-theory K3 geometry
degenerates into the union of two rational elliptic surfaces meeting
along the heterotic $T^{2}$ \cite{Morrison:1996pp}.

The other parameter is the heterotic string coupling, which corresponds
to K3 volume on the M-theory side, with weak heterotic coupling corresponding
to zero volume for the K3 surface. Going to this limit takes us out
of the regime where 11D supergravity is a reliable approximation to
M-theory, but because we are considering highly supersymmetric compactifications,
the duality results are expected to persist when we add M2-brane effects.
Again, there is an analogous limit in 8D F/het duality: in that case,
the heterotic coupling is dual to the area of a section of the elliptic
fibration, which may be interpreted as the area of the base of the
fibration \cite{Morrison:1996pp}. 

\subsection{Duality in 4D}

By fibering the 7D M/heterotic duality adiabatically over an $S^{3}$
base, we should be able to obtain dual pairs that give the same 4D
effective theory. From the M-theory side, for this theory to have
$N=1$ SUSY, the total space of the K3 fibration must have holonomy
$G_{2}$. Additionally, we want to look at effective theories with
non-abelian gauge symmetry, so that our space will be a $G_{2}$ orbifold.
In the large heterotic volume limit, the heterotic geometry is determined
to be a suborbifold of the $G_{2}$ orbifold, and SUSY then requires
that it is an SYZ fibration of a Calabi--Yau orbifold (i.e. a special
Lagrangian $T^{3}$ fibration of such a space over an $S^{3}$ base)
\cite{Acharya:1996ci}. The topology of $G_{2}$ and Calabi--Yau orbifolds
requires that our fibrations have singular fibers (by which we mean
fibers with multiple components in their resolution) where the adiabatic assumption
will break down\footnote{Because of this violation of the adiabatic assumption, it is not guaranteed
that the duality results will persist in 4D. In the notation of \cite{Sen:1996na},
our case is of type 2(b), where duality often persists despite the
presence of singular fibers.}. Such fibrations of $G_2$ manifolds were considered in an adiabatic limit in \cite{donaldson2017adiabatic}.

The large-volume limits on the heterotic side of the duality requires
all geometric radii to be large compared to the relevant dimensionful
parameter, which sets up a hierarchy of scales: we require that the
$T^{3}$ fibers are large compared to $\left(\alpha'\right)^{3/2}$
but small compared to the volume of the base\footnote{In our torus-orbifold setup, volumes are to be interpreted as products
of radii in the torus covering space. }. On the M-theory side, the K3 fibers on the $G_{2}$ side are also
required to be small compared to the volume of the base. 

For our 4D duality, we will apply the half-K3 and weak coupling limits
fiberwise. This means that we will work in a corner of the $G_{2}$
moduli space where each K3 fiber, including the singular fibers, grows
a long throat and simultaneously shrinks to small volume. This fiberwise
half-K3 limit translates to a ``half-$G_{2}$'' limit, where our
$G_{2}$ space grows a long throat with a Calabi--Yau threefold fiber that degenerates at the ends. The duality in this
limit identifies the generic Calabi--Yau fiber as the heterotic
geometry. By introducing a fibration, we also introduce additional
possibilities for configurations of singularities in our half-$G_{2}$
compared to our half-K3. We will restrict ourselves to orbifold (i.e.
codimension four) singularities, which live along a three-dimensional
locus. These loci may be confined to the endpoints of the throat interval,
in which case we will have a similar picture to the half-K3 limit,
but they also may stretch across the throat interval and intersect
the generic Calabi--Yau fiber. In the latter case, the singularities
are higher codimension in the two boundary fibers and give rise to non-perturbative
effects from the perspective of the heterotic compactification. 

\subsection{F-Theory Duals}

A useful tool in analyzing the heterotic string and M-theory
has been duality with F-theory, so this could be a candidate
to use in a search for an algorithmic construction of heterotic duals
to given M-theory backgrounds, as was done in \cite{Braun:2017uku}.
However, in our case, where we are looking at isolated points of enhanced
gauge symmetry in moduli space, the fiberwise nature of the data and
the complex structures required by the dualities prevent a straightforward
implementation of this method. 

To see the limitation, consider an M-theory background on a K3-fibered
$G_{2}$ manifold. If we apply the 7D M/heterotic duality, we obtain
bundle and flat connection data on the $T^{3}$ fibers of the heterotic
geometry $Y$, i.e. the duality gives the restrictions $E\bigm|_{A}$
of the heterotic gauge bundle $E$ to each $T^{3}$ fiber $A\subset Y$.
This by itself is not enough information to reconstruct $E$---we
have the vertical data but not the horizontal data. In the case of
an elliptic fibration, where the vertical data is given by a spectral
cover, the horizontal data is provided by a line bundle over that
spectral cover \cite{Friedman:1997yq}. 

In the case of M/heterotic duality, the $T^{3}$-fibration of $Y$
is a special Lagrangian fibration, which requires a choice of complex
structure where the holomorphic coordinates are made by pairing real
coordinates on the base and on the fiber. This means that there is
no elliptic curve contained in the $T^{3}$ fibers, and therefore
we do not have bundle data on any elliptic fibration of $Y$. Thus
an F-theory dual of the heterotic model cannot be used to infer the
missing bundle data. The F-theory dual can be constructed only
after we are able to determine the bundle by other means. 

The complex structure change that would be required for an application of an F-theory
dual may be thought of in $N=2$ language as a movement in the hypermultiplet
moduli space. In the case of a generic heterotic gauge bundle, where
one would be moving from one generic point of the moduli space to
another, an F-theory dual may give the correct answer (although even
this generic situation may be complicated by the presence of domain
walls in the moduli space). However, our situation deals with non-generic
bundles with point-like instantons on orbifold singularities, and
a shift in the hypermultiplet moduli space is likely to change the
matter spectrum, especially because the bundle moduli of fractional-holonomy
point-like instantons are coupled to the geometric moduli of the singular
spaces on which they reside \cite{Aspinwall:1997ye}. 

\section{M-Theory on Joyce Orbifolds}

Now we will describe the M-theory backgrounds for which we would like
to find candidate heterotic duals. For the purposes of this paper,
we will think of low-energy M-theory as 11D supergravity supplemented
by 7D spectra from M2 branes, as in \cite{Anderson:2006mv}. Then,
an M-theory compactification is specified by a choice of background metric,
C-field, and 7D gauge field. Here
we will consider $G_{2}$ orbifolds $X$ of the form $T^{7}/\Gamma$,
where $\Gamma$ is a finite group, and we will assume vanishing C-field
and gauge field backgrounds\footnote{While background C-field flux on a smooth $G_{2}$ manifold necessarily
breaks supersymmetry \cite{Acharya:2000ps}, some $G_{2}$ orbifolds
can support background C-field fluxes and gauge fields at the singular
loci that together preserve supersymmetry \cite{Acharya:2002kv}.
It would be interesting to investigate heterotic duals of these cases. }. 

The non-abelian factors in the gauge group of the low-energy effective
theory may be read off from the locus $S$ of orbifold singularities
in $X$, which comes from the fixed points of elements of $\Gamma$.
Each connected component of the orbifold locus of codimension four gives rise to gauge symmetry
in the effective theory according to the ADE classification of the
singularity \cite{Witten:1995ex}. In the examples we consider, each component of the singular locus is topologically $T^{3}$ or $T^{3}/\mathbb{Z}_{2}$. Counting these components on
the M-theory side gives the non-abelian gauge symmetry of the low
energy theory. The gauge group will have an additional abelian factor
$\text{U}(1)^{b_{\Gamma}^{2}(X)}$ from the Kaluza-Klein reduction
of the M-theory C-field, where $b_{\Gamma}^{2}(X)$ counts the number
of $\Gamma$-invariant harmonic 2-forms on $T^{7}$. Isometries of
the metric give an additional low-energy abelian gauge symmetry of dimension
$b_{\Gamma}^{1}(X)$. In our $N=1$ supersymmetric cases, we have $b_{\Gamma}^{1}(X)=0$
and $b_{\Gamma}^{2}(X)=0$, so that the 4D low-energy gauge group
has no abelian factor. 

In addition to gauge bosons, the massless spectrum of M-theory on
$X$ includes chiral multiplets that may or may not be charged under
the gauge symmetry. The number of uncharged chiral multiplets is determined
by $b_{\Gamma}^{3}(X)$, the number of $\Gamma$-invariant harmonic
3-forms on $X$. The charged matter, meanwhile, is determined by the
geometry of the orbifold loci: each codimension four locus component  
$L$ contributes $b^{1}(L)$
chiral multiplets valued in the adjoint of the gauge group factor
corresponding to $L$ \cite{Acharya:1998pm}. Intersections of the
orbifold loci give rise to more complicated matter representations,
but the examples considered in this paper have non-intersecting loci,
so will be limited to adjoint matter. All of the matter in our examples
lies in real representations of the gauge group, so the spectra are
non-chiral. 

Because gauge symmetry and charged matter in the low-energy theory
is specified by the orbifold singularities of $X$, it is independent
of a choice of K3 fibration. However, to compare this spectrum to
that of a dual heterotic string, we must choose a particular K3 fibration
$\pi:X\to Q$ and relate the gauge theory of the 4D effective theory
to that of the 7D effective theories on the fibers. For example, the
$\text{SU}(2)^{16}$ gauge symmetry on a generic $T^{4}/\mathbb{Z}_{2}$
fiber will be reduced to a subgroup in the 4D theory because the relevant
components of the orbifold locus intersect the generic fiber at multiple
points, so that these singularities appear to be distinct from the
perspective of the theory on the fiber, but not from the perspective
of $X$. In other words, the monodromy action of $\Gamma$ on the
singularities of the fiber reduces the gauge group to a subgroup in
4D. 

\subsection{Examples}

Now we will discuss details of three M-theory backgrounds that will
serve as our examples for which we will identify candidate heterotic
duals in the half-$G_{2}$ limit. Our $G_{2}$ orbifolds are of the form $T^{7}/\mathbb{Z}_{2}^{3}$, where $\mathbb{Z}_{2}^{3}$ is generated
by elements $\alpha,\beta,$ and $\gamma$. All three examples have
the same actions for $\alpha$ and $\beta$ on $T^{7}$ but differ
in the action of $\gamma$. The first two generators act as
\begin{eqnarray*}
\alpha:(x_{1},...,x_{7}) & \mapsto & (-x_{1},-x_{2},-x_{3},-x_{4},x_{5},x_{6},x_{7})\\
\beta:(x_{1},...,x_{7}) & \mapsto & (-x_{1},\tfrac{1}{2}-x_{2},x_{3},x_{4},-x_{5},-x_{6},x_{7}) \ ,
\end{eqnarray*}
where each $x_i \sim x_i + 1$ is a coordinate on a circle. 
Each of these elements fixes 16 $T^{3}$'s in $T^{7}$, while exchanging
the fixed tori of the other element in pairs. The element $\alpha\beta$
acts freely on $T^{7}$. Quotienting $T^{7}$ by the action of $\Gamma_{1}=\left\langle \alpha,\beta\right\rangle $
gives the $G_{2}$ orbifold 
\[
X_{1}=T^{7}/\Gamma_{1}\cong \left( T_{123456}^{6}/\left\langle \alpha,\beta\right\rangle \right) \times S_{7}^{1} \ ,
\]
where subscripts on tori denote their coordinates. At
this stage, the orbifold does not have full holonomy $G_{2}$, and
will preserve $N=2$ SUSY in 4D, as discussed in the first example
below. 

The 6-orbifold factor in $X_{1}$ is an orbifold limit of a Borcea--Voisin
Calabi--Yau threefold with Hodge numbers $(19,19)$ known as the Schoen
manifold\footnote{This orbifold may also be referred to as DW(0-2) \cite{Donagi:2008xy,Nibbelink:2012de} }.
We will discuss this orbifold further in section 4, where it serves
as the heterotic geometry in our $N=1$ examples. 

For our M-theory backgrounds, we will quotient the space $X_{1}$
further by an action of $\gamma$. In our first example, the action
of $\gamma$ is trivial and $N=2$ SUSY is preserved in 4D, while
the remaining examples have nontrivial $\gamma$ and preserve $N=1$
SUSY in 4D. 

\begin{table}
\begin{tabular}{|>{\centering}p{0.15\linewidth}|>{\centering}p{0.25\linewidth}|>{\centering}p{0.2\linewidth}|>{\centering}p{0.3\linewidth}|}
\hline 
Example Number & $\gamma$ Action & Low-Energy Gauge Symmetry & Massless Charged Matter 

($N=1$ Language)\tabularnewline
\hline 
\hline 
3.1 & Trivial & $\text{SU}(2)^{16}\times\text{U}(1)^{4}$ & \noindent \centering{}3 adjoint chirals per $\text{SU}(2)$\tabularnewline
\hline 
3.2 & Includes shift on $x_{3}$ & $\text{SU}(2)^{12}$ & 3 adjoint chirals per $\text{SU}(2)$\tabularnewline
\hline 
3.3 & No shift on $x_{3}$ & $\text{SU}(2)^{8}\times\text{SU}(2)^{8}$ & 3 adjoint chirals for $8$ $\text{SU}(2)$ factors and $1$ adjoint chiral
for other $8$ $\text{SU}(2)$ factors \tabularnewline
\hline 
\end{tabular}
\noindent \centering{}\caption{Summary of spectra of M-Theory backgrounds}
\end{table}

\subsubsection*{Example 3.1: $N=2$ SUSY}

First, we will consider the case where the action of $\gamma$ is
trivial, so that we are compactifying M-theory on the orbifold $X_{1}=T^{7}/\Gamma_{1}$
above. Ultimately, we are interested in $N=1$ SUSY in 4D, where the
orbifolds have full holonomy $G_{2}$, but non-perturbative features
of the half-$G_{2}$ limit appear in this simpler situation as well,
so it will serve as our first example. 

The space $X_{1}$ has 16 $T^{3}$'s of $A_{1}$ singularities, with
8 coming from $\alpha$ and 8 coming from $\beta$. Its orbifold Betti
numbers, by which we mean the counts of independent $\Gamma_{1}$-invariant
harmonic forms, are $b_{\Gamma_{1}}^{1}=1,\ b_{\Gamma_{1}}^{2}=3,$
and $b_{\Gamma_{1}}^{3}=11$. Thus, the gauge symmetry of the 4D theory
is expected to be $\text{SU}(2)^{16}\times\text{U}(1)^{4}$. The massless
matter spectrum is $3$ adjoint chirals of each $\text{SU}(2)$ plus
$11$ neutral chiral multiplets, where the count of adjoint chirals comes from $b^1(T^3) = 3$.

There are two immediate coassociative fibrations by Kummer orbifolds:
\begin{itemize}
\item The $\alpha$-fibration $\pi_{567}:T^{7}/\Gamma_{1}\to T_{567}^{3}/\left\langle \beta\right\rangle $
with generic fiber $T_{1234}^{4}/\left\langle \alpha\right\rangle $ 
\item The $\beta$-fibration $\pi_{347}:T^{7}/\Gamma_{1}\to T_{347}^{3}/\left\langle \alpha\right\rangle $
with generic fiber $T_{1256}^{4}/\left\langle \beta\right\rangle $ 
\end{itemize}
Given a choice of the $F$-fibration, where $F$ is one of $\alpha$
or $\beta$, let $Q_{1,F}$ be the 3-orbifold base of the fibration.
In this case, both $Q_{1,\alpha}$ and $Q_{1,\beta}$ are orbifold-equivalent
to $S^{1}\times P$, where $P$ is the pillow 2-orbifold obtained
as the quotient of $T^{2}$ by a reflection in both coordinates. Topologically,
this base is $S^{2}\times S^{1}$, and it has four non-linking circles
of singularities. 

Each of these fibrations will determine a dual heterotic model. In
either case, we want to take the base orbifold to be large compared
to both the fiber and the scale set by the gravitational coupling
$\kappa$, meaning in particular that the $S_{7}^{1}$ factor is large.
We are thus in the limit of a strongly-coupled IIA model on $T_{123456}^{6}/\Gamma_{1}$.
By moving in the geometric moduli space to small $S_{7}^{1}$, and
thus small IIA coupling, one may apply additional tools of IIA/het
duality, but it is possible that the adiabatic assumption is violated
in this limit. See section 7 for more discussion of Type IIA duals. 

\begin{figure}
\begin{centering}
\includegraphics[scale=0.25]{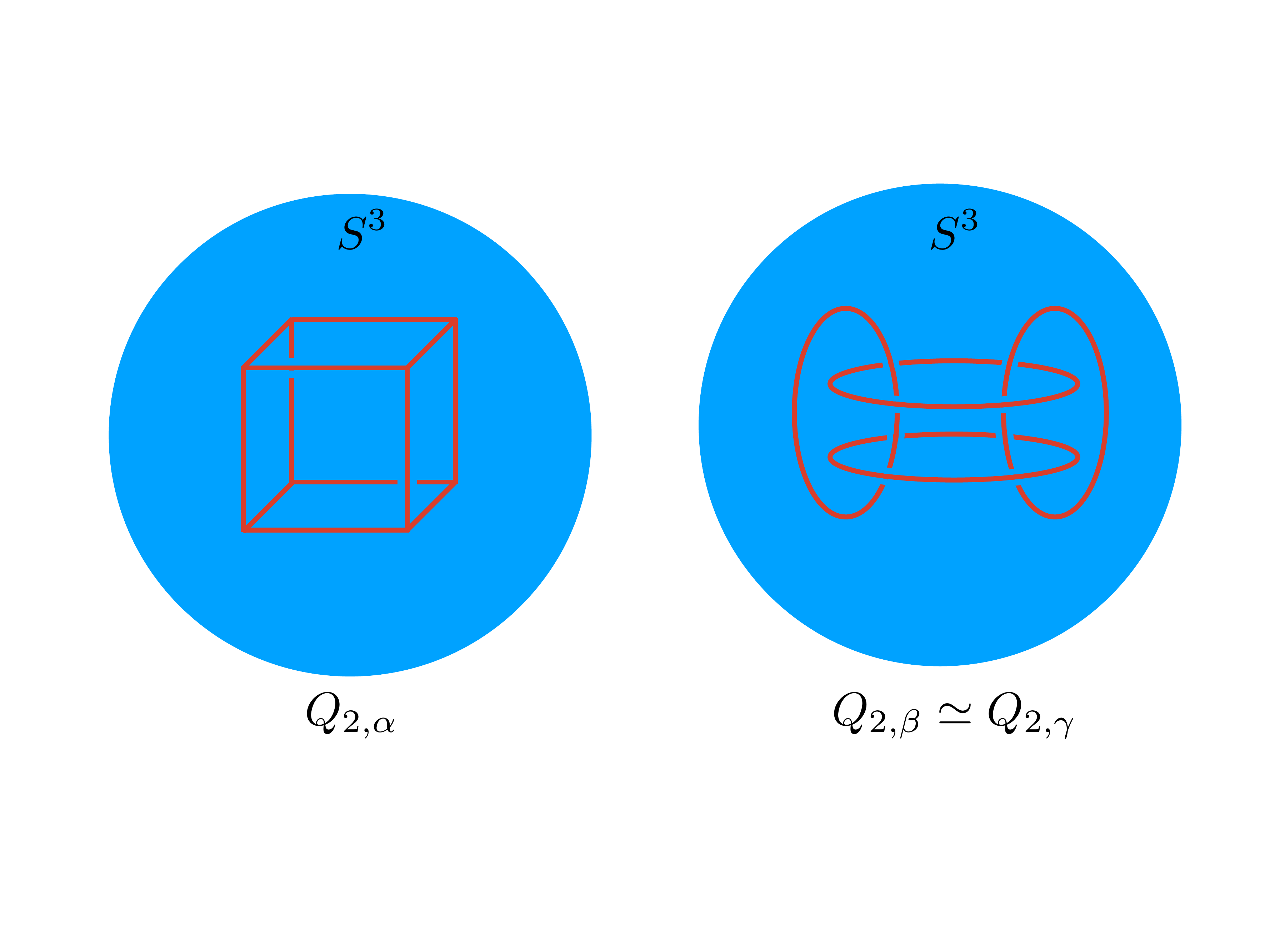}
\par\end{centering}
\caption{The base 3-orbifolds for the $\alpha$, $\beta$, and $\gamma$ fibrations
of the $G_{2}$ orbifold $X_{2}$. In all cases, the base orbifold
is of the form $T^{3}/\mathbb{Z}_{2}^{2}$, and is homeomorphic to
a 3-sphere. There is a 1-dimensional locus of singularities that in
the case of the $\alpha$-fibration is the 1-skeleton of a cube, while
in the $\beta$- and $\gamma$-fibrations it is a doubled Hopf link.
These orbifolds serve as the bases for the fibrations of $X_{3}$
as well, with $Q_{3,\alpha}\simeq Q_{3,\beta}\simeq Q_{2\alpha}$
and $Q_{3,\gamma}\simeq Q_{2,\beta}\simeq Q_{2,\gamma}$. The dual
heterotic geometries are $T^{3}$ fibrations over the same bases. }
\end{figure}

\subsubsection*{Example 3.2: The Simplest Joyce Orbifold}

Next, let us move on to examples that preserve $N=1$ SUSY in 4D.
First, we will consider the Joyce orbifold defined by the third generator
\[
\gamma_{2}:(x_{1},...,x_{7})\mapsto(\tfrac{1}{2}-x_{1},x_{2},\tfrac{1}{2}-x_{3},x_{4},-x_{5},x_{6},-x_{7})\ .
\]
This orbifold was first considered in \cite{10.4310/jdg/1214458109}
and studied further in \cite{Liu:1998tha}. Let $\Gamma_{2}\cong\left\langle \alpha,\beta,\gamma_{2}\right\rangle $
and $X_{2}=T^{7}/\Gamma_{2}$. In this case, the actions of $\alpha,\beta,$
and $\gamma_{2}$ are symmetric: $\gamma_{2}$ fixes 16 $T^{3}$'s in
$T^{7}$, just as $\alpha$ and $\beta$ do, and it acts freely on
the fixed loci of the other elements, as they do on the 16 $T^{3}$'s
fixed by $\gamma_{2}$. Altogether, we find $12$ $T^{3}$'s of $A_{1}$
singularities ($4$ from each of $\alpha,\beta,$ and $\gamma_{2}$).
The orbifold Betti numbers in this case are $b_{\Gamma_{2}}^{1}=0,\ b_{\Gamma_{2}}^{2}=0,$
and $b_{\Gamma_{2}}^{3}=7$. Thus in the low energy theory we expect
$\text{SU}(2)^{12}$ gauge symmetry with $3$ adjoint chirals for
each $\text{SU}(2)$ and $7$ neutral chiral multiplets.

In addition to the two coassociative Kummer fibrations inherited from
$X_{1}$, the orbifold $X_{2}$ has an additional fibration coming
from the action of $\gamma_{2}$. These three fibrations are:
\begin{itemize}
\item The $\alpha$-fibration $\pi_{567}:T^{7}/\Gamma_{2}\to T_{567}^{3}/\left\langle \beta,\gamma_{2}\right\rangle $
with generic fiber $T_{1234}^{4}/\left\langle \alpha\right\rangle $ 
\item The $\beta$-fibration $\pi_{347}:T^{7}/\Gamma_{2}\to T_{347}^{3}/\left\langle \alpha,\gamma_{2}\right\rangle $
with generic fiber $T_{1256}^{4}/\left\langle \beta\right\rangle $ 
\item The $\gamma_{2}$-fibration $\pi_{246}:T^{7}/\Gamma_{2}\to T_{246}^{3}/\left\langle \alpha,\beta\right\rangle $
with generic fiber $T_{1357}^{4}/\left\langle \gamma_{2}\right\rangle $ 
\end{itemize}
Given a choice of the $F$-fibration, where $F$ is one of $\alpha,\beta,$
or $\gamma_{2}$, we let $H_{2,F}\cong\mathbb{Z}_{2}^{2}$ be the
group generated by the two generators of $\Gamma_{2}$ other than
$F$, and we let $Q_{2,F}$ be the 3-orbifold base of the fibration,
which is topologically $S^{3}$ in all cases. In each case, $H_{2,F}$
will act trivially on one of the 7 coordinates. This is the coordinate
that should be chosen as the throat direction in the half-$G_{2}$
limit. 

Now, let us examine the $\alpha$-fibration of $X_{2}$, following
example 3.1 of \cite{Liu:1998tha}. We will discuss this first example
of a $N=1$ fibration in detail and be more brief in subsequent examples.
The action of $H_{2,\alpha}$ on $T_{567}^{3}$ has the fixed point
loci 
\begin{eqnarray*}
\text{Fix}(\pi_{567}\circ\beta) & = & \left\{ x_{5}\in\left\{ 0,\tfrac{1}{2}\right\} ,x_{6}\in\left\{ 0,\tfrac{1}{2}\right\} \right\} \\
\text{Fix}(\pi_{567}\circ\gamma_{2}) & = & \left\{ x_{5}\in\left\{ 0,\tfrac{1}{2}\right\} ,x_{7}\in\left\{ 0,\tfrac{1}{2}\right\} \right\} \\
\text{Fix}(\pi_{567}\circ\beta\gamma_{2}) & = & \left\{ x_{6}\in\left\{ 0,\tfrac{1}{2}\right\} ,x_{7}\in\left\{ 0,\tfrac{1}{2}\right\} \right\} \ , 
\end{eqnarray*}
which are each 4 disjoint circles. We have 
\[
\#\left[\text{Fix}(\pi_{567}\circ\beta)\cap\text{Fix}(\pi_{567}\circ\gamma_{2})\cap\text{Fix}(\pi_{567}\circ\beta\gamma_{2})\right]=8 \ , 
\]
and these 8 points of intersection are the only elements in the intersection
of any two of these loci. Because any intersection of the loci involves
three circles, and these circles become line intervals $S^{1}/\mathbb{Z}_{2}$
under the $H_{2,\alpha}$ quotient, the elements in the intersection
correspond to trivalent vertices in the graph of fixed points on the
base; the graph is the 1-skeleton of a cube (see Figure 2). Denote
the base orbifold $T_{567}/H_{2,\alpha}$ by $Q_{2,\alpha}$ and its
orbifold locus by $\Sigma_{Q_{2,\alpha}}$. 

Let us examine how the singular locus of $X$ lies with respect to
the $\alpha$-fibration. The four components that come from fixed
$T^{3}$ of $\alpha$ become 4 disjoint multi-sections of $\pi_{567}$,
so that they provide the 16 $A_{1}$ singularities in each Kummer
fiber. The remainder of the singular locus lies over $\Sigma_{Q_{2,\alpha}}$. The components coming from fixed $T^{3}$ of $\beta$ project under
$\pi_{567}$ to the edges of $\Sigma_{Q_{2,\alpha}}$ parallel to
the $x_{7}$ axis, while the components from $\gamma_{2}$ project
onto edges parallel to the $x_{6}$ axis.

The singular fibers (by which we mean fibers that have multiple components in their resolution) of the $\alpha$-fibration are those that lie above
above $\Sigma_{Q_{2,\alpha}}$. The fibers that project to an edge
of $\Sigma_{2,Q_{\alpha}}$ are acted upon by one element of $H_{2,\alpha}$ and have multiplicity 2. The fibers lying above a corner of $\Sigma_{Q_{2,\alpha}}$ are acted upon by all of $H_{2,\alpha}$ and have multiplicity 4. Note that $H_{2,\alpha}$ acts trivially on $x_{4}$, so that this
should be our choice of K3 throat coordinate in this case.

If we consider instead the $\beta$-fibration, we find similar results
but with a different base orbifold $Q_{\beta}$. In this case, the
relevant fixed point loci are
\begin{eqnarray*}
\text{Fix}(\pi_{347}\circ\alpha) & = & \left\{ x_{3}\in\left\{ 0,\tfrac{1}{2}\right\} ,x_{4}\in\left\{ 0,\tfrac{1}{2}\right\} \right\} \\
\text{Fix}(\pi_{347}\circ\gamma) & = & \left\{ x_{5}\in\left\{ \tfrac{1}{4},\tfrac{3}{4}\right\} ,x_{7}\in\left\{ 0,\tfrac{1}{2}\right\} \right\} \\
\text{Fix}(\pi_{347}\circ\alpha\gamma) & = & \varnothing \ .
\end{eqnarray*}
This gives us the orbifold locus $\Sigma_{Q_{2,\beta}}$ that is four
disjoint circles forming a doubled Hopf link. See example 3.2 of \cite{Liu:1998tha}
for details. In contrast to the cube locus of the $\alpha$-fibration,
the locus $\Sigma_{\beta}$ has no vertices, so that the singular
fibers are of multiplicity 2 only. This makes the monodromy analysis
somewhat simpler in the heterotic dual theory. Finally, the $\gamma_{2}$-fibration
gives results identical to the $\beta$-fibration up to change of
coordinates. 

\subsubsection*{Example 3.3: Orbifold Singular Loci}

Our second $N=1$ background is similar to the previous example, except
for a shift in the action of $\gamma$. This time we define the third
group generator 
\[
\gamma_{3}:(x_{1},...,x_{7})\mapsto(\tfrac{1}{2}-x_{1},x_{2},-x_{3},x_{4},-x_{5},x_{6},-x_{7}) \ , 
\]
which is identical to $\gamma_{2}$ except for the lack of shift on
$x_{3}$. The orbifold defined by this choice of third generator was
studied in \cite{10.4310/jdg/1214458110} and used for M-theory compactification
in \cite{Acharya:1996ci}. The element $\gamma_{3}$ still fixes $16$
$T^{3}$'s in $T^{7}$, but now $\left\langle \alpha,\beta\right\rangle $
does not act freely on these \textbf{$16$} $T^{3}$'s and instead
orbifolds them to $8$ $T^{3}/\mathbb{Z}_{2}$'s. The action of $\alpha\beta$
kills two of the harmonic 1-forms on $T^{3}$, so that $b_{\left\langle \alpha\beta\right\rangle }^{1}\left(T_{246}^{3}/\left\langle \alpha\beta\right\rangle \right)=1$.
This modifies the spectrum of massless charged matter. 

As before, define $\Gamma_{3}=\left\langle \alpha,\beta,\gamma_{3}\right\rangle $
and $X_{3}=T^{7}/\Gamma_{3}$. The Betti numbers of $X_{3}$ are identical
to those of $X_{2}$, since the shifts on the coordinates do not affect
the harmonic forms. The singular loci of $X_{3}$ are $8$ $T^{3}$
and $8$ $T^{3}/\mathbb{Z}_{2}$ of $A_{1}$ singularities. Thus,
we expect low-energy gauge symmetry $\text{SU}(2)^{16}$, with $3$
adjoint chiral multiplets each for $8$ of these $\text{SU}(2)$ factors
and $1$ adjoint chiral multiplet each for the remaining $\text{SU}(2)$ factors.
Additionally, there will be $7$ neutral chiral multiplets, as in
example 3.2. 

The coassociative Kummer fibrations are defined in the same way for
this example as for example 3.2. The difference is that the base of
the $\beta$-fibration has changed. The singular loci $\Sigma_{Q_{3,\alpha}}$
and $\Sigma_{Q_{3,\beta}}$ are the 1-skeleton of a cube, as was $\Sigma_{Q_{2,\alpha}}$,
while the singular locus $\Sigma_{Q_{3,\gamma_{3}}}$ is the doubled
Hopf link, as was $\Sigma_{Q_{2,\gamma_{2}}}$. 

\section{The Dual Heterotic Geometry}

Given a $G_{2}$ orbifold $X=T^{7}/\Gamma$ with a choice of K3 fibration,
we want to identify the dual Calabi--Yau orbifold $Y$ on which to
compactify the heterotic string. To obtain $Y$, we replace the K3
fibers of $X$ by dual $T^{3}$ fibers with metric determined by the
K3 data. Because we want large heterotic volume, we work in the half-$G_{2}$
limit on the M-theory side, where the heterotic geometry is given
by the generic fiber transverse to the throat direction. The complex
structure on the heterotic orbifold may be determined by demanding
that the orbifold group act holomorphically on $T^{6}$, and this
gives a complex structure compatible with the SYZ condition, which
requires that the $T^{3}$ fibers are special Lagrangian. Different
choices of K3 fibration on the M-theory side give rise to different
heterotic geometries, but they are biholomorphic; all of our $N=1$
examples give orbifold limits of the Schoen manifold \cite{Liu:1998tha},
similar to the results of \cite{Braun:2017uku} for twisted-connected
sums. However, the $T^{3}$ fibrations of these biholomorphic spaces
are inequivalent, and in particular they have bases with topologically
distinct singular loci, as we saw for the K3 fibrations of the $G_{2}$
orbifolds in section 3. 

As the heterotic geometry is a fiber of the $G_{2}$ orbifold,
it intersects the singular loci of the ambient space. In particular,
in our examples, each $T^{3}$ singular locus of the $G_{2}$ orbifold
intersects the heterotic geometry either trivially or in two disconnected
$T^{2}$. (A helpful lower-dimensional picture is to imagine $T^{2}$
as a $S^{1}$-fibration over an interval that is branched at the two
endpoints.) Thus when we have only $T^{3}$ singular loci in the $G_{2}$
orbifold, the number of components of the heterotic singular locus
is twice the number of components of the M-theory singular locus that
lie parallel to the throat coordinate. The $T^{3}/\mathbb{Z}_{2}$
loci, on the other hand, intersect the heterotic geometry either trivially
or in only one $T^{2}$, so there is no doubling of loci. The singular
loci in the heterotic geometry are expected to give rise to non-perturbative
gauge symmetry when they carry point-like instantons, as we will discuss
in detail in the following sections. 

In the remainder of this section, we will describe the heterotic geometries
dual to the examples 3.1, 3.2, and 3.3 that we introduced in the previous
section. 

\subsubsection*{Example 4.1: $N=2$ SUSY }

In the $N=2$ case of example 3.1, the $\alpha$- and $\beta$-fibrations
are equivalent up to a change of coordinates, so we may study the
dual geometry from either perspective. For definiteness, we will choose
the $\alpha$-fibration. Both $x_{3}$ and $x_{4}$ fit our criteria
for the throat coordinate and give biholomorphic results, so we choose
$x_{4}$ as the throat coordinate, as this is the option that will
survive the further $\gamma$-action of the $N=1$ examples. This
means that we stretch the $x_{4}$ direction and look at our $G_{2}$
space as a fibration $\pi_{4}:X_{1}\to S^{1}/\left\langle \alpha\right\rangle $
over the resulting long interval $S^{1}/\left\langle \alpha\right\rangle \cong\left[0,\tfrac{1}{2}\right]$.
The fiber above a point away from the ends of the interval is our
dual geometry $Y_{1,\alpha}=T_{123567}^{6}/\left\langle \beta\right\rangle $.
(Note that the action of $\alpha$ only descends to the fibers at
$x_{4}=0,\frac{1}{2}$. Away from these points, it serves only to switch the 6-orbifold
fiber with an identical ``far away'' fiber.)

The space $Y_{1,\alpha}$ is constructed as a fibration $\pi_{567}:Y_{1,\alpha}\to Q_{1,\alpha}$
over the same base 3-orbifold $Q_{1,\alpha}$ as on the M-theory side,
but with the generic Kummer fiber $T_{1234}^{4}/\left\langle \alpha\right\rangle $
replaced by a flat 3-torus $T_{123}^{3}$ and with holonomies around
the singular fibers determined by those on the M-theory side. The
Betti numbers of our space are found to be 
\[
b_{\left\langle \beta\right\rangle }^{1}\left(Y_{1,\alpha}\right)=2,\quad b_{\left\langle \beta\right\rangle }^{2}\left(Y_{1,\alpha}\right)=7,\quad b_{\left\langle \beta\right\rangle }^{3}\left(Y_{1,\alpha}\right)=12\ .
\]

The complex structure of $Y_{1,\alpha}$ is constrained by the SYZ
condition and the holomorphy of the action of $\beta$, but, unlike
in the $N=1$ cases below, this information is not enough to fully
determine the complex structure---there is an $S^{2}$ of complex
structures compatible with these conditions. 

For another perspective on this space, we may rewrite it as $T_{123567}^{6}/\left\langle \beta\right\rangle \cong \left(T_{1256}^{4}/\left\langle \beta\right\rangle \right) \times T_{37}^{2}$,
so we have a trivial fibration of Kummer orbifolds $T^{4}/\mathbb{Z}_{2}$
over $T^{2}$. From this perspective, we see that the space has $16$
$T^{2}$'s of $A_{1}$ singularities, corresponding to $T^{2}\sqcup T^{2}$
cross sections of the $8$ $T^{3}$ singular loci of the M-theory
geometry that come from $\beta$. When projected to the base, the
singular $T^{2}$'s project to the singular $S^{1}$'s of $Q_{1,\alpha}$
in groups of four. 

\begin{center}
\begin{figure}
\begin{centering}
\includegraphics[scale=0.45]{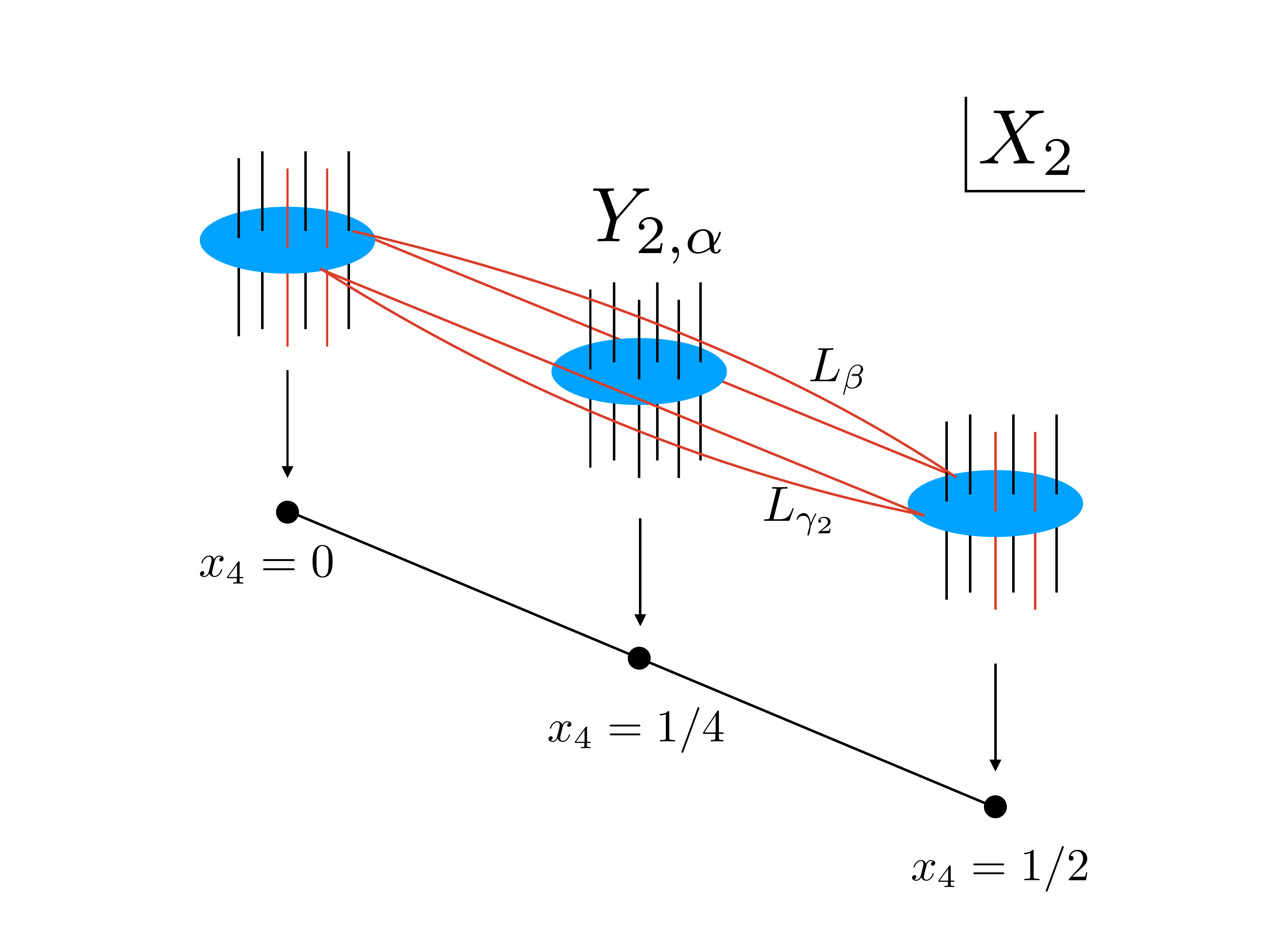}
\par\end{centering}
\caption{A schematic view of the half-$G_{2}$ limit of the $G_{2}$ orbifold
$X_{2}$ from example 3.2 with the $\alpha$-fibration. We have stretched
$X_{2}$ along the direction of $x_{4}$, the throat coordinate. The
heterotic dual geometry $Y_{2,\alpha}$ is the inverse image $\pi_{4}^{-1}\left(\frac{1}{4}\right)$,
and is shown with its SYZ fibration of $T^{3}$ fibers (black lines)
over the 3-orbifold base $Q_{2,\alpha}$ (blue disk). Some of the
black lines are singular fibers that do not create singularities in
the total space; the singularities in the total space are displayed
by red lines. The $\alpha$-fixed loci (vertical red lines) are confined
to the ends of the $x_{4}$ interval, while the $\beta$-fixed loci
$L_{\beta}$ and $\gamma_{2}$-fixed loci $L_{\gamma_{2}}$ stretch
across the interval. These $T^{3}$ loci that stretch across the interval
intersect $Y_{2,\alpha}$ in a 2-component locus $T^{2}\sqcup T^{2}$.
The monodromy action of $\alpha$ on the singular $T^{2}$ of $Y_{2,\alpha}$
fixed by $\beta$ is to travel around a loop in $x_{4}$ that begins
at $x_{4}=\frac{1}{4}$, passes through $x_{4}=0$ or $x_{4}=\frac{1}{2}$,
and returns to $x_{4}=\frac{1}{4}$ along the other leg of $L_{\beta}$,
so that the singular $T^{2}$'s are swapped in pairs. }

\end{figure}
\par\end{center}

\subsubsection*{Example 4.2: Duals to Fibrations of $X_{2}$ }

Next, we will examine the dual geometries to fibrations of $X_{2}$,
studied in example 3.2 above. We will begin with the $\alpha$-fibration,
which is similar to our previous example, but with an additional $\mathbb{Z}_{2}$
action by $\gamma_{2}$ (see Figure 3). Because $\gamma_{2}$ acts
nontrivially on $x_{3}$, the only coordinate of $T^{7}$ that can
act as the throat coordinate of the half-$G_{2}$ limit is $x_{4}$,
so the relevant fibration for this limit is $\pi_{4}:X_{2}\to S_{4}^{1}/\left\langle \alpha\right\rangle $,
where $S_{4}^{1}$ is taken to be large. The fiber above a point away
from the ends of the interval is our dual geometry $Y_{2,\alpha}=T_{123567}^{6}/H_{2,\alpha}$,
where, as in example 3.2, $H_{2,\alpha}=\left\langle \beta,\gamma_{2}\right\rangle $.
The $T^{3}$ fibration dual to the $\alpha$-fibration of $X$ is
$\pi_{567}:Y_{2,\alpha}\to Q_{2,\alpha}$, with generic fiber $T_{123}^{3}$.
Then $\pi_{567}$ is an SYZ fibration of the Borcea--Voisin Calabi--Yau
orbifold $Y_{2,\alpha}$. 

The Betti numbers of this example are 
\[
b_{H_{2,\alpha}}^{1}\left(Y_{2,\alpha}\right)=0,\ b_{H_{2,\alpha}}^{2}\left(Y_{2,\alpha}\right)=3,\ b_{H_{2,\alpha}}^{3}\left(Y_{2,\alpha}\right)=8 \ , 
\]
and these will be the same for our remaining $N=1$ heterotic geometries,
which are all homeomorphic. 

To see that $Y_{\alpha}$ is a Borcea--Voisin orbifold, we note that
$\beta$ acts nontrivially only on the $1256$ coordinates, and $T_{1256}^{4}/\left\langle \beta\right\rangle $
is a Kummer surface. Furthermore, $\gamma_{2}$ acts as $(-1)$ on
the holomorphic 2-form $dz_{1}\wedge dz_{2}$ of the Kummer surface,
and if we shift the coordinate on the remaining torus $T_{37}$ to
be $w_{3}=z_{3}-\frac{i}{4}$, then $\gamma_{2}$ acts as $w_{3}\mapsto-w_{3}$,
as required.

Because we want an SYZ fibration by the $T_{123}^{3}$ fibers, the
complex structure must pair fiber and base coordinates. Additionally,
we demand that $H_{2,\alpha}$ acts holomorphically, and this leaves
a unique choice of complex structure: 
\begin{eqnarray*}
z_{1} & = & ix_{1}+x_{5}\\
z_{2} & = & ix_{2}+x_{6}\\
z_{3} & = & ix_{3}+x_{7}
\end{eqnarray*}
so that our projection map $\pi_{567}:T_{123567}^{6}\to T_{567}^{3}$
is 
\[
\begin{pmatrix}z_{1}\\
z_{2}\\
z_{3}
\end{pmatrix}\mapsto\text{Re}\begin{pmatrix}z_{1}\\
z_{2}\\
z_{3}
\end{pmatrix}
\]
and our group $H_{2,\alpha}$ acts as 
\begin{eqnarray*}
\beta:(z_{1},z_{2},z_{3}) & \mapsto & \left(-z_{1},\frac{i}{2}-z_{2},z_{3}\right)\\
\gamma:(z_{1},z_{2},z_{3}) & \mapsto & \left(\frac{i}{2}-z_{1},z_{2},\frac{i}{2}-z_{3}\right)\\
\beta\gamma:(z_{1},z_{2},z_{3}) & \mapsto & \left(z_{1}-\frac{i}{2},\frac{i}{2}-z_{2},\frac{i}{2}-z_{3}\right) \ .
\end{eqnarray*}
Furthermore, if we restrict $\alpha$ to the heterotic geometry, we
find the involution 
\[
\alpha\bigm|_{Y_{2,\alpha}}:(z_{1},z_{2},z_{3})\mapsto\left(\overline{z}_{1},\overline{z}_{2},\overline{z}_{3}\right) \ , 
\]
so in the 7D space, $\alpha$ acts as a complex conjugation map between
$Y_{2,\alpha}$ and a distant fiber. 

The singularities in our threefold are the fixed point loci 
\begin{eqnarray*}
\text{Fix}(\beta) & = & \left\{ x_{1}\in\left\{ 0,\tfrac{1}{2}\right\} ,x_{2}\in\left\{ \tfrac{1}{4},\tfrac{3}{4}\right\} ,x_{5}\in\left\{ 0,\tfrac{1}{2}\right\} ,x_{6}\in\left\{ 0,\tfrac{1}{2}\right\} \right\} \\
\text{Fix}(\gamma_{2}) & = & \left\{ x_{1}\in\left\{ \tfrac{1}{4},\tfrac{3}{4}\right\} ,x_{3}\in\left\{ \tfrac{1}{4},\tfrac{3}{4}\right\} ,x_{5}\in\left\{ 0,\tfrac{1}{2}\right\} ,x_{7}\in\left\{ 0,\tfrac{1}{2}\right\} \right\} \\
\text{Fix}(\beta\gamma_{2}) & = & \varnothing \ .
\end{eqnarray*}
The first two loci are each 16 disjoint complex curves with $\text{Fix}(\beta)\cap\text{Fix}(\gamma_{2})=\varnothing$.
The action of $\beta$ on $\text{Fix}(\gamma_{2})$ identifies the
curves in pairs, as does the action of $\gamma_{2}$ on $\text{Fix}(\beta$),
so we will have 16 curves of $A_{1}$ singularities in $Y_{2,\alpha}$. 

Different choices of K3 fibration on the M-theory side give rise to
heterotic orbifolds that are biholomorphic, but may have different
metrics (determined by the radii of the covering $T^{6}$) and different
SYZ fibrations. To illustrate this, we will look at the heterotic
geometry dual to the $\beta$-fibration of $X_{2}$. The throat coordinate
must now be chosen as $x_{6}$, because this is the coordinate that
is inverted by $\beta$ while being fixed by $H_{2,\beta}=\text{\ensuremath{\left\langle \alpha,\gamma\right\rangle }}$.
Thus we take $S_{6}^{1}$ to be large and the heterotic geometry $Y_{2,\beta}$
will be realized as the generic fiber of $\pi_{6}:X_{2}\to S_{6}^{1}/\left\langle \beta\right\rangle $.
This space is again an SYZ fibration with generic fiber $T^{3}$,
but this time over the base $Q_{2,\beta}$, which we saw in example
$3.2$ is inequivalent to $Q_{2,\alpha}$, since the singular locus
of the former is a doubled Hopf link, while the singular locus of
the latter is the 1-skeleton of a cube. Despite the change in base,
the total space $Y_{2,\beta}=T_{123457}^{6}/H_{2,\beta}$ with the
complex structure determined by SYZ and $H_{2,\beta}$ is biholomorphic
to $Y_{2,\alpha}$. Additionally, the heterotic geometry $Y_{2,\gamma_{2}}=T_{123456}^{6}/H_{2,\gamma_{2}}$
that results from the choice of the $\gamma_{2}$-fibration is biholomorphic
to the first two examples and has an SYZ fibration equivalent to that
of $Y_{2,\beta}$.

Thus, the choice of fibration of $X_{2}$ only affects the metric
on the dual heterotic geometry. Because our M/heterotic duality requires
a particular geometric limit where the throat direction is stretched
and the base of the SYZ fibration is much larger than its fibers,
a change in K3 fibration on the M-theory side requires a change of
metric on the heterotic side to ensure the correct cycles are large
or small. In our torus orbifold cases, this only requires a rescaling
of the radii of the covering torus. We will see in the next example
that the choice of fibration has other important effects on the heterotic
gauge bundle. 

\subsubsection*{Example 4.3: Dual Geometries for Orbifold Singular Loci}

Finally, let us look at heterotic dual geometries for $X_{3}$, which
has singular loci homeomorphic to the nonsingular orbifold $T^{3}/\mathbb{Z}_{2}$.
Despite this change, we find that the heterotic geometry is again biholomorphic
to the one found in example 4.3 for all choices of fibrations. 

We begin with the $\alpha$-fibration, which is similar to the $\alpha$-fibration
of example 4.3 except for the configuration of the singular loci.
Our geometry in this case is $Y_{3,\alpha}=T_{123567}^{6}/H_{3,\alpha}$,
where $H_{3,\alpha}=\left\langle \beta,\gamma_{3}\right\rangle $.
The fixed loci of $T^{6}$ in this case are 
\begin{align*}
\text{Fix}(\beta) & =\left\{ x_{1}\in\left\{ 0,\tfrac{1}{2}\right\} ,x_{2}\in\left\{ \tfrac{1}{4},\tfrac{3}{4}\right\} ,x_{5}\in\left\{ 0,\tfrac{1}{2}\right\} ,x_{6}\in\left\{ 0,\tfrac{1}{2}\right\} \right\} \\
\text{Fix}(\gamma_{3}) & =\left\{ x_{1}\in\left\{ \tfrac{1}{4},\tfrac{3}{4}\right\} ,x_{3}\in\left\{ 0,\tfrac{1}{2}\right\} ,x_{5}\in\left\{ 0,\tfrac{1}{2}\right\} ,x_{7}\in\left\{ 0,\tfrac{1}{2}\right\} \right\} \\
\text{Fix}(\beta\gamma_{3}) & =\varnothing \ ,
\end{align*}
where the only change relative to the previous example is the $x_{3}$
coordinate of the $\gamma_{3}$-loci. As before, each of $\beta$
and $\gamma_{3}$ acts on the fixed loci of the other to reduce the
number of components by a factor of 2. Thus, we again find a Calabi--Yau
orbifold of the form $T^{6}/\mathbb{Z}_{2}^{2}$ with 16 $A_{1}$
singularities. The $8$ $T^{2}$'s in the $\gamma_{3}$-fixed loci of
$Y_{3,\alpha}$ are $T^{2}$ cross-sections of the $T^{3}/\mathbb{Z}_{2}$
loci in the ambient $G_{2}$ orbifold. Note that the $\mathbb{Z}_{2}$
action does not descend to the $T^{2}$'s in $Y_{3,\alpha}$ because
it is accomplished by the element $\alpha\beta\in\Gamma_{3}$, which
inverts the $x_{4}$ coordinate and thus exchanges $Y_{3,\alpha}$
with a different fiber of the half-$G_{2}$ limit. 

The $\beta$-fibration gives identical results to the $\alpha$-fibration
(unlike in example 4.2), and the $\gamma_{3}$-fibration gives identical
results to that of the $\gamma_{2}$-fibration of example 4.3. Thus,
all of our $N=1$ fibration examples have biholomorphic heterotic
geometries. This is not surprising in light of the results of \cite{Braun:2017uku},
where it was found that all smooth TCS $G_{2}$ backgrounds have heterotic duals based on the same Schoen Calabi--Yau. The complexity of heterotic compactifications
come from the choices of gauge bundles, and indeed we will see in
section 6 that the heterotic duals of the $\alpha$- and $\gamma$-fibrations
of example 3.3 have different instanton configurations. 

\section{The Heterotic Gauge Bundle}

Now we move on to the more subtle part of the heterotic background:
the gauge bundle\footnote{Because we are working with orbifolds, we are really constructing
gauge \emph{sheaves} or \emph{orbibundles}, but we will continue to
informally use the word ``bundle'' for these objects.}. The information necessary to construct this bundle is contained
in the data of the M-theory metric, C-field background, and 7D gauge
field background. Given a K3 fibration of a $G_{2}$ manifold, we
may apply 7D M/heterotic duality to each fiber to find the restriction
of the heterotic gauge bundle to each dual $T^{3}$ fiber. 

Ideally, the restriction of the bundle to each $T^{3}$ fiber, along
with the monodromies around the singular fibers, would allow us to
reconstruct the gauge bundle over the entire Calabi--Yau space. In
the case of an elliptic fibration of a Calabi--Yau manifold, the work
of \cite{Friedman:1997yq} allows one to do exactly that. However,
their methods rely on the fact that the elliptic curve is a complex
manifold, so their results are not so easily generalized to $T^{3}$
fibers. As described in section 2.4, part of the data required for
the gauge bundle reconstruction of \cite{Friedman:1997yq} is a choice
of line bundle over a spectral cover which corresponds in F-theory
to an instanton bundle on the background D7-branes. The analogous
data in an M-theory compactification would seem to be a background
instanton configuration for the gauge theories living on the singular
loci, but such backgrounds have not been thoroughly studied. 

Reconstructing the bundle in general cases may be possible with better
understanding of the special Lagrangian structure of the fibers within
the Calabi--Yau, but we do not yet have the tools to work with this
data. For now, we will study the gauge bundle from the perspective
of the point-like instantons required to cancel anomalies. These instantons
give rise to non-perturbative gauge symmetry and matter, and we may
attempt to match their spectra with the M-theory side. Insight into
instanton behavior may also be found from dual Type I models, where
D5-branes play the role of the dual object \cite{Berkooz:1996iz,Aldazabal:1997wi,Aldazabal:1998mr}. 

There are at least three levels of checks one may perform to give
evidence for a conjectured dual pair:
\begin{enumerate}
\item The most coarse check is to ensure that the two sides give the same
effective 4D gauge symmetry. In the case of point-like instantons,
we may refine this criterion by splitting the gauge symmetry into
a perturbative and non-perturbative part from the heterotic perspective,
and checking that each part of the gauge symmetry matches with what
is given on the M-theory side. 
\item Next, one can check that the massless charged matter agrees on the
two sides. For point-like instantons on orbifold singularities, the
massless spectrum is well-understood only in simple examples. 
\item A third level to check is that the low energy effective action agrees
on the two sides of the duality. Unfortunately, the action associated
to excitations about point-like instantons on orbifold singularities
has not been investigated, so there are not currently quantitative
checks to be made. However, one can reason qualitatively about the
action by considering which modes should be massive or massless at
specific points in moduli space. 
\end{enumerate}
In this paper, we will focus primarily on the coarsest check: the
gauge symmetry of the low-energy effective theory. We will start by
describing the split between heterotic perturbative and non-perturbative
spectra and reviewing some results about spectra of point-like instantons
on orbifold singularities. 

\subsection{Perturbative vs. Non-Perturbative Spectra}

Although we work in the weak heterotic string coupling limit $\lambda\to0$
where possible, anomaly cancellation guarantees that near the singular
loci of our heterotic geometry, the background will exhibit phenomena
that are non-perturbative in the string coupling, such as point-like
instantons. Thus the massless spectrum from the heterotic string is
best understood as a sum of a perturbative part (the spectrum seen
by a 2D CFT description) and a non-perturbative part, which cannot
be seen from the CFT perspective. This approach was refined in heterotic
orbifold compactifications in \cite{Aldazabal:1997wi}, where it was
argued that because the string worldsheet perspective cannot describe
the non-perturbative part of the massless spectrum, the perturbative
spectrum is no longer constrained by modular invariance. Instead,
the requirement is that the combined perturbative and non-perturbative
spectra have no anomalies in the low-energy effective theory.

Relevant examples of perturbative spectra may be constructed from non-singular instantons on orbifold loci. A basic configuration is the $\text{SU}(2)$-instanton on $\mathbb{R}^4/\mathbb{Z}_2$ described in \cite{Berkooz:1996iz}, which is obtained as a $\mathbb{Z}_2$-quotient of the standard $\text{SU}(2)$-instanton configuration with $c_2=1$ centered at the origin of $\mathbb{R}^{4}$. If we write $\text{SO}(4)=\left(\text{SU}(2)_{L}\times\text{SU}(2)_{R}\right)/\mathbb{Z}_{2}$ and embed the gauge group $\text{SU}(2)$ as either $\text{SU}(2)_{L}$ or $\text{SU}(2)_{R}$, the resulting $\text{SO}(4)$-connection has a monodromy $M$ on the lens space $S^{3}/\mathbb{Z}_{2}$ at infinity given by $M=-I_{4}$, where $I_{4}$ is the rank-4 identity matrix. Denote this connection on $\mathbb{R}^{4}/\mathbb{Z}_{2}$ by ${\cal A}_0$. We will use this type of instanton in Section 6 to build non-singular bundle configurations on our heterotic orbifolds that reproduce the perturbative spectra seen in our dual M-theory models. When these instantons shrink to zero size, they produce additional effects, as we will discuss in the next subsection. Similar non-singular instantons may be built by starting with calorons, instantons on $\mathbb{R}^3 \times S^1$ periodic up to a gauge transformation \cite{Lee:1998vu,Kraan:1998pm}. These configurations are made of constituent BPS monopoles and are naturally centered at pairs of points, making them more relevant to the examples at hand. 

For M/heterotic duality in 7 non-compact dimensions, the entire spectrum
is visible perturbatively in the half-K3 limit, since the moduli space
of M-theory on K3 coincides with that of the perturbative heterotic
string on $T^{3}$. When this duality is fibered over a 3D base, we
expect the singular fibers to introduce phenomena that are non-perturbative
from the heterotic side. We can identify the effects that come from
singular fibers by the same geometric criterion that is used in heterotic/F-theory
duality \cite{Bershadsky:1996nh}: the gauge symmetry and matter that
come from components of the singular locus that meet the generic K3
fiber transversely should be visible perturbatively on the heterotic
side, while that coming from components that project to nonzero codimension
on the base should come from mechanisms that are invisible to perturbation
theory\footnote{Note that this rule applies only to matter from singular loci that are codimension-four in the total space, as in our examples. Codimension-seven loci, for instance, give perturbative matter while projecting to nonzero codimension on the base}. An alternative characterization used in IIA/heterotic duality
is that degenerate K3 fibers on the IIA side that require multiple
components in their resolution correspond to non-perturbative effects
on the heterotic side \cite{Braun:2016sks}. 

The perturbative dictionary tells us that the data for an $E_{8}$
bundle on $T^{3}$ is stored in the choice of a half-K3 surface whose
boundary is the given $T^{3}$. This is analogous to Looijenga's theorem
that the data for an $E_{8}$ bundle on an elliptic curve is contained
in an embedding of the curve into a $k=8$ del Pezzo surface \cite{looijenga1976root,looijenga1980invariant}.
Meanwhile, the non-perturbative part of the gauge symmetry will come
from point-like instantons sitting on orbifold singularities. Singular gauge bundles coming from point-like instantons on orbifold singularities
are not fully understood or classified, but we will review some of
what is known. 

\subsection{Point-Like Instantons on Orbifold Singularities}

In our flat orbifold examples, the inclusion of point-like instantons
is required by the heterotic anomaly cancellation condition:
\[
dH=\alpha'\left(\text{tr}F\wedge F-\text{tr}R\wedge R\right) \ ,
\]
which for $dH=0$ forces a gauge bundle for which the second Chern
character (i.e. the Poincare dual of the homology class of the instanton
distribution) agrees with that of the tangent sheaf of the orbifold
(at least in a formal sense). In other words, we are forced to place
instantons along the orbifold loci. In the dimensions transverse to
the loci, these look like point-like instantons. The right-hand side of the anomaly cancellation condition may be modified non-perturbatively by the presence of background NS5-branes. We work in a limit where any wrapped NS5-branes are represented by point-like instantons \cite{Choi:2019ovk}, so that both perturbative and non-perturbative contributions are contained in the $\text{tr}F\wedge F$ term.

This type of configuration is further motivated by the supersymmetry conditions:
because we are working in the half-K3 limit, $\alpha'$ corrections
are suppressed, and the supersymmetry condition requires that we have
a Hermitian-Yang-Mills connection on our bundle. This condition, in
combination with anomaly cancellation, requires the connection to
be flat away from the singular loci, while on these loci it has instanton
number matching the background metric. To see this, we write the anomaly
cancellation condition as $\text{tr}F\wedge F=0$ and the SUSY D-term
equation as $\star F=-\omega\wedge F$, where $\omega$ is the K{\"a}hler
form. Wedging $F$ with both sides and then taking a trace gives us
\[
\text{tr}\left(F\wedge\star F\right)=-\omega\wedge\text{tr}\left(F\wedge F\right)=0 \ .
\]
The left hand side is the norm-squared of the gauge field strength,
so it vanishes away from orbifold loci. Together, these conditions
tell us that we must place point-like instantons on our orbifold loci,
and that there is no freedom to vary the connection away from these
loci other than choosing holonomies. It is possible that the gauge
fields could have nontrivial profiles along the singular loci, but
because we chose a trivial background configuration for the 7D gauge
fields on the M-theory side, we expect the profiles to be trivial
on the heterotic side as well. 

In our $N=1$ examples, we have additional constraints on the gauge
bundles that arise from the properties of the massless spectrum calculated
from M-theory:
\begin{enumerate}
\item There is no abelian gauge symmetry in the 4D effective theory, meaning
no tensor multiplets in a local 6D description near a singular locus. 
\item All charged matter in 4D is in the adjoint representation. Because
point-like instantons typically come with fundamental multiplets,
this suggests that there may be Higgsing of the non-perturbative spectrum. 
\end{enumerate}
With these points in mind, we can look at the effects of point-like
instantons on the massless spectrum. A point-like instanton comes
with extra massless particles that are non-perturbative in the string
coupling. There are several ways to understand this phenomenon: one
can think of it as a stringy ``smoothing'' of an apparent geometric
singularity via extra massless particles, or as the massless sector
of the worldvolume theory of a wrapped NS5-brane or a wrapped M5-brane
in a dual theory, or as a theory of tensionless strings. Point-like
instantons behave differently in the $E_{8}\times E_{8}$ and $\text{Spin}(32)/\mathbb{Z}_{2}$
heterotic theories. Because our primary duality gives an $E_{8}\times E_{8}$
model, one may expect that only $E_{8}\times E_{8}$ point-like instantons
are relevant. However, the instantons in our backgrounds behave like
T-dual $\text{Spin}(32)/\mathbb{Z}_{2}$ instantons, similar to cases
examined in \cite{Berkooz:1996iz,Aldazabal:1997wi}.

First let us briefly review what happens when you shrink $E_{8}$
point-like instantons to zero size on a smooth 6D geometry \cite{Seiberg:1996vs,Ganor:1996mu}.
Because this case isn't directly relevant to us, we will just summarize
the spectrum: on a smooth point, an $E_{8}$ point-like instanton
gives rise to an extra massless tensor and no extra gauge symmetry.
From the point of view of heterotic-M theory, with M-theory compactified
on $Y\times S^{1}/\mathbb{Z}_{2}$, where $Y$ is a Calabi--Yau threefold,
a point-like instanton may be thought of as an M5-brane wrapped on
$Y$ that moves from the interior of the interval to the boundary
\cite{Ovrut:2000qi}. In this picture, the VEV of the scalar in the
tensor multiplet controls the position of the M5-brane along the interval. 

Note that in this case and in later cases, the extra massless
particles can be blocked by the presence of a nontrivial B-field holonomy
on the orbifold point \cite{Aspinwall:1998he}. Indeed, to fully specify
a heterotic dual, we must choose a background of B-field holonomies
on the 2-cycles of our space. The holonomies on the $T^{3}$ fibers
are determined by the shape of the K3 fibers of the $G_{2}$ orbifold,
as shown in \cite{Lu:1998xt} by matching moduli. There can be no
holonomies on the base, as it is homeomorphic to $S^{3}$, but there
may be B-field holonomies with one leg along a fiber and one leg along
the base. This case includes the singular loci as well as any extra
2-cycles of the space. 

In our examples, the point-like instantons reside on orbifold points
of the geometry. Because this is a worse bundle singularity than the
point-like instantons on a smooth point, extra non-perturbative multiplets
can arise \cite{Aspinwall:1997ye,Aspinwall:1998xj,Aspinwall:1998he,Intriligator:1997dh}.
For point-like instantons on an orbifold point, the holonomy of the
gauge bundle may be nontrivial, since the lens space surrounding the
orbifold point has nontrivial fundamental group. The case with trivial
holonomy was investigated in \cite{Aspinwall:1997ye}. In \cite{Aspinwall:1998xj},
simple cases of nontrivial holonomy were worked out. It was established
in \cite{Aspinwall:1998he} that an $E_{8}\times E_{8}$ point-like
instanton with nontrivial holonomy on an orbifold point does not give
rise to a tensor multiplet, but retains its non-perturbative gauge
symmetry and charged matter. This can be understood from the heterotic-M
theory perspective, where a wrapped M5-brane cannot move from the
orbifold point into the bulk because it must preserve its holonomy.
Thus a point-like instanton with nontrivial holonomy may be thought
of as a frozen singularity in the bundle. In some cases, this may
be interpreted in terms of fractional M5-branes \cite{Tachikawa:2015wka}. 

In the cases considered in this paper, the orbifold singularities
of the heterotic geometry look locally like an $A_{1}$ singularity
$\mathbb{C}^{2}/\mathbb{Z}_{2}$, so we will review options for fractional
$E_{8}\times E_{8}$ instantons on such a space, following section
4.3 of \cite{Aspinwall:2000fd}. The only nontrivial option for the
holonomy is $\mathbb{Z}_{2}$, and there are two ways that this may
be embedded in $E_{8}$, up to conjugacy:
\begin{enumerate}
\item It may be embedded so as to have centralizer $\left(E_{7}\times\text{SU}(2)\right)/\mathbb{Z}_{2}$.
This gives instanton number $c_{2}=1/2$ and no tensor multiplet nor
gauge symmetry. 
\item It may be embedded so as to have centralizer $\text{Spin}(16)/\mathbb{Z}_{2}$.
This gives $c_{2}=1$ and a non-perturbative $\text{SU}(2)$, but no
tensor. 
\end{enumerate}
We may combine these types of instantons to get new examples. For
instance, we may place both a trivial holonomy instanton and the $c_{2}=1/2$
instanton on an $A_{1}$ singularity to get an instanton with $c_{2}=3/2$
that gives no tensor multiplet, but a non-perturbative $\text{SU}(2)$
so that the gauge symmetry in the visible sector becomes $E_{7}\times\text{SU}(2)$.
This is the situation that corresponds to the tangent sheaf of $\mathbb{C}^{2}/\mathbb{Z}_{2}$. 

What kinds of instantons are allowed when there are multiple singularities?
The case of the tangent sheaf of $T^{4}/\mathbb{Z}_{2}$, which has
16 $A_{1}$ singularities, is discussed in \cite{Aspinwall:1998he,Ludeling:2014oba}
and has the behavior of $16$ independent instantons, each with $c_{2}=3/2$.
The behavior of the heterotic backgrounds in our examples suggests
that there exist also configurations where the instantons residing
on different loci are not independent. In other words, we seem to
have instantons that are only semi-localized, so that they spread
their instanton number evenly over two loci. In the case of an instanton
semi-localized on an $A_{1}\oplus A_{1}$ singularity, the resulting
non-perturbative gauge symmetry is only $\text{SU}(2)$. The gauge
fields localized on the two singularities must take values in the
diagonal $\mathfrak{su}(2)$ subalgebra of the $\mathfrak{su}(2)\oplus\mathfrak{su}(2)$
that would arise from separate instantons on the two loci. A compactification
on $T^{4}/\mathbb{Z}_{2}$ with $8$ such semi-localized instantons
suggests that each one has instanton number $c_{2}=3$, the sum of
the instanton numbers for each locus. One candidate for these instantons is the singular limit of a $\mathbb{Z}_2$-quotient of an $\text{SU}(2)$ caloron. 

While our main duality relates M-theory to the $E_{8}\times E_{8}$
heterotic string, we will also be interested in an alternate duality
to the $\text{Spin}(32)/\mathbb{Z}_{2}$ string. This dual model involves
point-like instantons as well, so we will review some properties of
this case. The $\text{Spin}(32)/\mathbb{Z}_{2}$ point-like instantons
behave oppositely to the $E_{8}\times E_{8}$ ones with respect to
their spectrum: they produce non-perturbative vector multiplets when
placed on a smooth point, and augment these with tensor multiplets
when placed on orbifold singularities \cite{Aspinwall:1997ye,Blum:1997mm}.
There are multiple types of $\text{Spin}(32)/\mathbb{Z}_{2}$ instantons,
but we are interested in particular in those that live on $\mathbb{Z}_{2}$
orbifold singularities and participate in the duality with Type I
on $T^{4}/\mathbb{Z}_{2}$ \cite{Bianchi:1990tb,Gimon:1996rq,Berkooz:1996iz}.
In the case that on the Type I side distributes one half-D5-brane
at each fixed point, the heterotic background carries a combination
of two point-like instantons at each fixed point. Each points has
a ``hidden'' $c_{2}=1$ instanton with no low-energy gauge symmetry
or tensor multiplets. On top of this background, there is a configuration
of fractional D5-branes, which may also be interpreted as point-like
instantons. When the D5-branes are distributed evenly across the fixed
points, and in the absence of Wilson lines, the gauge group is $\text{SU}(16)\times\text{U}(1)$,
where a rank 16 factor has been removed by a Green-Schwarz-type mechanism
\cite{Berkooz:1996iz}.

\subsection{Point-Like Instanton Spectra }

Ideally, we would be able to verify that the spectra of our heterotic
backgrounds agree with those of their purported M-theory duals. This
goal is hampered by the fact that calculating spectra of point-like
instantons on orbifold singularities is challenging and still not
fully understood in the literature. Existing results are generally
based on 6D anomaly cancellation (e.g. \cite{Intriligator:1997dh,Aldazabal:1997wi})
or F-theory duals (e.g. \cite{Aspinwall:1998he,Ludeling:2014oba}).
A pattern seems to emerge that $E_{8}\times E_{8}$ point-like instantons
on orbifold singularities do not give rise to adjoint matter; their
charged matter appears to be fundamental matter in all existing examples.
This provides a challenge for matching such spectra to those of M-theory
on our $G_{2}$ orbifolds, because the latter have only adjoint matter.
The semi-localized instantons suggested in the previous section, perhaps
combined with a Wilson line background, likely give rise to matter
valued in the adjoint of the diagonal subgroup. 

The spectrum of a heterotic orbifold with point-like instantons is
not limited to the non-perturbative spectrum of the instanton, but
also comprises a perturbative spectrum, split as usual into untwisted
and twisted sectors. A recipe for calculating the perturbative spectrum
is given in \cite{Aldazabal:1997wi}, where it is shown that an additional
energy term must be included in the left-moving twisted sector mass
formula to account for the magnetic flux of the instantons sitting
at the fixed point, thought of as wrapped M5-branes. In this paper,
we are interested in the non-perturbative gauge sector, so we leave
an investigation of the perturbative spectrum using this recipe for
future work.

One particularly relevant example appears in section 5 of \cite{Aldazabal:1997wi},
where anomaly cancellation in an $E_{8}\times E_{8}$ background on
$T^{4}/\mathbb{Z}_{3}$ is achieved by adding a non-perturbative $\text{SU}(2)^{9}$
factor to the gauge group along with charged hypermultiplets. This
is interpreted as a spectrum arising from frozen fivebranes in the
T-dual $\text{Spin}(32)/\mathbb{Z}_{2}$ theory. We will argue for
a similar interpretation of our non-perturbative gauge symmetry in
section 7. 

\section{Example Dual Pairs}

Equipped with preliminary analysis of the heterotic geometry and gauge
bundle, we now explore aspects of our candidate dual pairs. Because
we are primarily interested in the non-perturbative aspects of the
half-$G_{2}$ limit, we will give only a brief description of the
perturbative part of the analysis, but we include a construction method for non-singular instantons that replicate the perturbative spectra. 
We will begin with a description of the 7D duality shared by all three
examples and then discuss the details of each example individually. 

In all of our examples of M-theory on K3 fibrations, the generic fibers
are at the same $\mathbb{Z}_{2}$ orbifold point in K3 moduli space,
so they share the same effective 7D theory. In this case, the heterotic
dual background is a flat $T^{3}$ with three Wilson lines that branch
$E_{8}\times E_{8}$ to $\text{SU}(2)^{16}$ \cite{Harvey:1995ne,Braun:2009wh}.
The only non-gravitational supermultiplet in 7D is the vector multiplet,
so there is no charged matter from a 7D perspective. When further
compactified on $T^{3}$ to 4D, this perturbative spectrum becomes
$\text{SU}(2)^{16}$ gauge symmetry with $3$ adjoint chiral supermultiplets
for each $\text{SU}(2)$ (which is just the 4D $N=4$ vector multiplet
in 4D $N=1$ language). Additionally, there are abelian factors in
the gauge group as well as neutral chiral multiplets, but we will
ignore these parts of the spectrum, as they are not our primary interest.
In the following examples, we will use this 4D perturbative spectrum
as a starting point and add in the additional orbifold actions as
well as non-perturbative effects. 

\begin{table}
\begin{tabular}{|c|c|c|>{\centering}p{6cm}|}
\hline 
Example Number & Fibration & Perturbative Gauge Symmetry & Non-Perturbative Gauge Symmetry\tabularnewline
\hline 
\hline 
6.1 & $\alpha,\beta$ & $\text{SU}(2)^{8}\times\text{U}(1)^{4}$ & $\text{SU}(2)^{8}$\tabularnewline
\hline 
6.2 & $\alpha,\beta,\gamma$ & $\text{SU}(2)^{4}$ & $\text{SU}(2)^{8}$\tabularnewline
\hline 
6.3 & $\alpha,\beta$ & $\text{SU}(2)^{4}$ & $\text{SU}(2)^{12}$\tabularnewline
\hline 
6.3 & $\gamma$ & $\text{SU}(2)^{8}$ & $\text{SU}(2)^{8}$\tabularnewline
\hline 
\end{tabular}

\caption{Summary of gauge symmetry in heterotic duals}
\end{table}

\subsection{$N=2$ Example}

First, we will discuss the heterotic dual of the M-theory background
of example 3.1, which has a trivial action of $\gamma$. There are
$16$ disjoint $T^{3}$'s of $A_{1}$ singularities in the $G_{2}$
orbifold $X_{1}$, with $8$ coming from $\alpha$ and $8$ from $\beta$.
We saw that there are two choices of coassociative Kummer fibration
in this example, but they give equivalent heterotic dual geometries.
In either case, half of the singular loci of $X_{1}$ have a transverse
intersection with the generic fiber, meaning that we expect $\text{SU}(2)^{8}$
perturbative gauge symmetry and $\text{SU}(2)^{8}$ non-perturbative
gauge symmetry on the heterotic side. 

For definiteness, consider the $\alpha$-fibration, where we view
the M-theory geometry as a $T_{1234}^{4}/\left\langle \alpha\right\rangle $-fibration
over $T_{567}^{3}/\left\langle \beta\right\rangle $. In example 4.1,
we saw that the dual geometry in this case is a $T_{123}^{3}$-fibration
over the same base. We may write our heterotic geometry as the trivial
Kummer fibration $Y_{1}=T_{1256}^{4}/\left\langle \beta\right\rangle \times T_{37}^{2}$.
This space has $16$ disjoint $T^{2}$'s of $A_{1}$ singularities,
all from $\beta$. Note that the SYZ $T^{3}$ fibers are not fully
contained within the K3 fibers, so that the perturbative Wilson lines
along the $T^{3}$ fibers prevent the heterotic gauge bundle from
factorizing into a K3 component and a $T^{2}$ component, which complicates
potential applications of IIA/heterotic duality. 

From a perturbative orbifold perspective, we have the Wilson lines
described above on each $T_{123}^{3}$ fiber, and we also must determine
a $\mathbb{Z}_{2}$-action of $\beta$ on the perturbative heterotic
gauge bundle. We will assume that $\beta$ acts by the outer automorphism
that swaps the perturbative $E_{8}$ factors, as this is the gauge
bundle action that corresponds to the geometric origin of the gauge
symmetry on the M-theory side: in the $G_{2}$ orbifold, the action
of $\beta$ on the fixed loci of $\alpha$ is to swap them in pairs,
reducing the resulting non-perturbative gauge symmetry from $\text{SU}(2)^{16}$
to $\text{SU}(2)^{8}$. This agrees with the choice of the action
of $\beta$ on the heterotic gauge bundle, which will break to the
diagonal $E_{8}$, and branch this to $\text{SU}(2)^{8}$ when combined
with the Wilson lines. The adjoint chiral multiplets are identified
in pairs as well, leaving us with $3$ adjoint chirals for each $\text{SU}(2)$.

The non-perturbative part of the non-abelian spectrum is the same
as the perturbative part: an additional $\text{SU}(2)^{8}$ with $3$
adjoint chiral multiplets each. This part of the spectrum should come
from point-like instantons on the $\beta$-loci, meaning that we should
get $\text{SU}(2)^{8}$ gauge symmetry from $16$ $T^{2}$'s of $A_{1}$
singularities. This appears to be a puzzle, because there is nothing
to distinguish $8$ of the loci as those that produce gauge symmetry,
while the others do not. However, the loci are paired by the monodromy
action of $\alpha$ within the ambient space. We illustrate this with
an example (see Figure 4).

\begin{figure}
\begin{centering}
\includegraphics[scale=0.2]{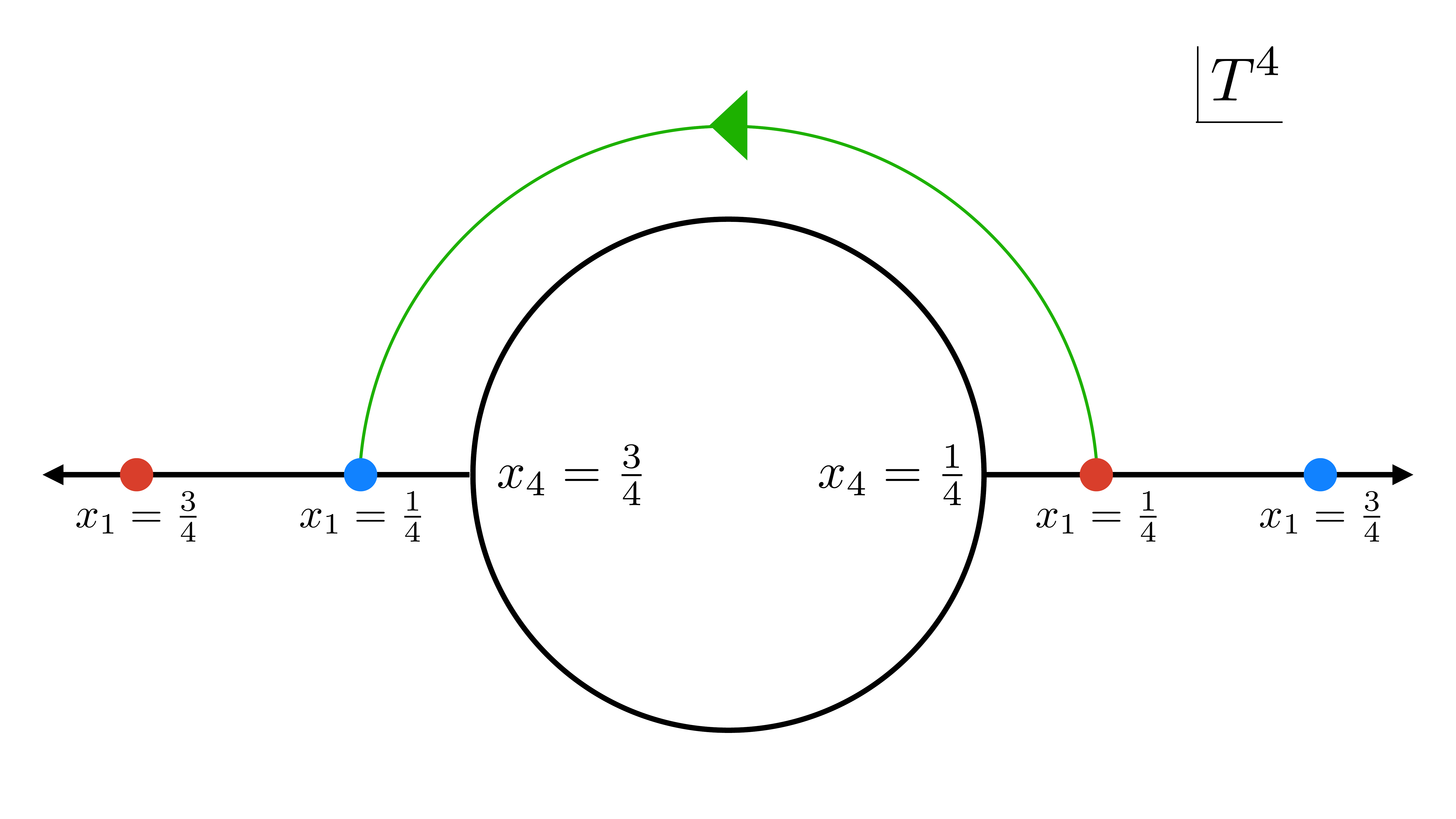}
\par\end{centering}
\caption{The action of $\alpha$-monodromy on a $T^{2}$ singular locus in
the $N=2$ example. Pictured is the $T^{4}$ within the covering $T^{7}$
that is defined by $x_{2}=x_{5}=x_{6}=0$. The $x_{3}$- and $x_{7}$-dimensions
are suppressed, so that each colored circle represents a $T^{2}$.
The fibers $\pi_{4}^{-1}\left(1/4\right)$ and $\pi_{4}^{-1}\left(3/4\right)$
are pictured, represented by the $x_{1}$-direction only. The two
$T^{2}$'s represented by red circles are interchanged by the action
of $\alpha$, as are those represented by blue circles. By following
the green contour from the $x_{4}=1/4$ fiber to the $x_{4}=3/4$
fiber and applying $\alpha$, one ends up with a monodromy action
by $\alpha$ on the singular loci of the $x_{4}=1/4$ fiber. }
\end{figure}

Within the heterotic geometry $Y_{1,\alpha}=\pi_{4}^{-1}\left(\frac{1}{4}\right)$,
consider the singular $T^{2}$ that is the image of $\left(\frac{1}{4},0,x_{3},\frac{1}{4},0,0,x_{7}\right)\subset T^{7}$,
where $x_{3}$ and $x_{7}$ are the $T^{2}$ coordinates. Suppose
we translate along the throat direction $x_{4}$ to a different Calabi--Yau
fiber located at $x_{4}=\frac{3}{4}$. Because our $T^{7}$ is identified
under the action of $\alpha$, which inverts the first four coordinates,
we have ended up back at $x_{4}=\frac{1}{4}$, and thus back within
$Y_{1,\alpha}$ at the point 
\[
\left(\frac{3}{4},0,-x_{3},\frac{1}{4},0,0,x_{7}\right) \ .
\]
If we perform this translation for every $(x_{3},x_{7})$, we obtain
a monodromy action by $\alpha$ that exchanges these two singular
$T^{2}$ within $Y_{1,\alpha}$. In general, this monodromy action
pairs up the $16$ singular $T^{2}$ of $Y_{1,\alpha}$. Our task
is to reproduce the effect of this geometric action within the heterotic
theory itself. The natural guess, given our constraints, is a semi-localized
instanton that is evenly distributed over the two $T^{2}$, as described
in section 5.2. This instanton ought to give rise to an $\text{SU}(2)$
gauge symmetry with three adjoint chiral multiplets (or, in $N=2$
language, an $\text{SU}(2)$ gauge symmetry with one adjoint hypermultiplet).
Thus we conjecture that the heterotic dual gauge bundle is comprised
of $8$ instantons of this type distributed across pairs of the singular
$T^{2}$ loci. This semi-localization may be understood from a T-dual
perspective as coming from a winding shift, as we will discuss in
the next section. 

Although the instanton is distributed over a disconnected locus, the separation is small because of the geometric limits
required for our duality with M-theory to be valid. The loci that
are paired by the instantons are separated only within the $T^{3}$
fiber, which is assumed to be small compared to the base for our duality
to hold, as described in section 2. In our example above, the two
singular $T^{2}$ both lie over $\left(0,0,x_{7}\right)$ in the base,
and their separation in the $x_{1}$-direction is infinitesimal compared
to the radius of $x_{7}$. On the other hand, the separation in the
$x_{1}$-direction is very large compared to $\sqrt{\alpha'}$, so
the disconnectedness demonstrated by this instanton is small compared to
the compactification volume, but large compared to the string scale.
The $\text{Spin}(32)/\mathbb{\mathbb{Z}}_{2}$ T-dual model of this
configuration is an asymmetric orbifold, as will be discussed below,
and thus a (weakly) non-geometric compactification. This non-geometric
aspect is not reflected in the geometry of the $E_{8}\times E_{8}$
model, but it leaves a remnant in the gauge
bundle. 

We may construct candidate configurations that reproduce the perturbative spectrum by deforming away from the point-like instanton limit and building a smooth instanton configuration on the orbifold $Y_1$ using copies of the connection ${\cal A}_0$ described in Section 5.1. We may use the monodromy $M=-I_4$, where $I_4$ denotes the rank-4 identity matrix, to match the Wilson line monodromies dictated by the half-K3 limit. We will work with the $\text{Spin}(32)/\mathbb{Z}_{2}$ string for convenience, but the procedure is similar for the $E_8 \times E_8$ string. Consider the triple of $\text{Spin}(32)/\mathbb{Z}_{2}$-monodromies
\begin{align*}
W_{1} & =\left(-I_{4},-I_{4},-I_{4},-I_{4},I_{4},I_{4},I_{4},I_{4}\right)\\
W_{2} & =\left(-I_{4},I_{4},-I_{4},I_{4},-I_{4},I_{4},-I_{4},I_{4}\right)\\
W_{3} & =\left(I_{4},-I_{4},I_{4},-I_{4},I_{4},-I_{4},I_{4},-I_{4}\right)\ ,
\end{align*}
where the notation indicates a block-diagonal matrix in $\text{Spin}(32)/\mathbb{Z}_{2}$. This triple breaks $\text{Spin}(32)/\mathbb{Z}_{2}\to\text{SO}(4)^{8}$. (In the case of the $E_{8}\times E_{8}$ string, we must instead replace $W_{1}$ by the Wilson line that breaks $E_{8}\to\text{SO}(16)$.) Let $A_{W}$ be the flat connection on $\left(T_{123}^{3}\times T_{567}^{3}\right)/\left\langle \beta\right\rangle$  that has monodromy $W_{i}$ along the $x_{i}$-direction for $i=1,2,3$. We will embed the $\text{SO}(4)$-instanton ${\cal A}_{0}$ into $\text{SO}(4)^{8}$ and place it at various fixed points of $T^{6}/\langle \beta \rangle$. Far from the fixed points, the instantons decay and match to the flat connection $A_{W}$. 

First, embed the connection ${\cal A}_{0}$ in the first four $\text{SU}(2)_{L}$
factors, and choose vanishing connections for all other $\text{SU}(2)$
factors of $\text{SO}(4)^{8}$. Denote this connection on $\mathbb{R}^{4}/\mathbb{Z}_{2}$
by
\[
{\cal A}_{1}=\left[\left(g_{L},1\right)\left(g_{L},1\right)\left(g_{L},1\right)\left(g_{L},1\right)\left(1,1\right)\left(1,1\right)\left(1,1\right)\left(1,1\right);W_{1}\right]\ ,
\]
where the notation indicates which components carry the instantons connections, and that the connection has monodromy $W_{1}$ around
the $x_{1}$ direction. This connection commutes locally with 
\[
\left(1,g_{R}\right)\left(1,g_{R}\right)\left(1,g_{R}\right)\left(1,g_{R}\right)\left(g_{L},g_{R}\right)\left(g_{L},g_{R}\right)\left(g_{L},g_{R}\right)\left(g_{L},g_{R}\right)\ ,
\]
which generates $\text{SU}(2)^{12}$. We place the connection ${\cal A}_{1}$
on a collection of the sixteen $T^{2}$ loci of $\mathbb{R}^{4}/\mathbb{Z}_{2}$
singularities to be discussed below. 

A similar connection ${\cal A}_{2}$ with monodromy $W_{2}$, to be supported
on a distinct set of four singular loci, is given by 
\[
{\cal A}_{2}=\left[\left(1,g_{R}\right)\left(1,1\right)\left(1,g_{R}\right)\left(1,1\right)\left(g_{L},1\right)\left(1,1\right)\left(g_{L},1\right)\left(1,1\right);W_{2}\right]\ .
\]
This connection commutes with a different $\text{SU}(2)^{12}$ such
that the sum of ${\cal A}_{1}$ and ${\cal A}_{2}$ gives a $\text{SO}(4)^{8}$-connection
whose centralizer is $\text{SU}(2)^{8}$, generated by 
\[
\left(1,1\right)\left(1,g_{R}\right)\left(1,1\right)\left(1,g_{R}\right)\left(1,g_{R}\right)\left(g_{L},g_{R}\right)\left(1,g_{R}\right)\left(g_{L},g_{R}\right)\ .
\]
Thus this instanton configuration reproduces the desired perturbative
gauge symmetry for the $N=2$ supersymmetric example. The matter spectrum
of the candidate instanton configuration is three adjoint chiral multiplets
per $\text{SU}(2)$ factor, as desired. These arise as the remaining freedom to choose flat connections for the unbroken $\text{SU}(2)$ factors: the six directions of the covering $T^6$ give six adjoints, which form three chiral multiplets. 

This method of building instanton configurations creates the correct perturbative spectrum, but it is not immediately clear how to place the summands ${\cal A}_1$ and ${\cal A}_2$ on the correct $T^2$ loci as dictated by the half-$G_2$ limit. In the point-like limit, we expect a $\mathbb{Z}_2$ symmetry such that every $\text{SU}(2)$-instanton is associated to a pair of $T^2$ loci. However, placing separate ${\cal A}_0$ instantons on these loci does not give the correct counting of $c_2$. The instanton configuration that behaves appropriately in the point-like limit likely begins with an instanton on $\left(\mathbb{R}^3\times S^1\right)/\mathbb{Z}_2$ that does not arise from local $\mathbb{R}^4/\mathbb{Z}_2$ instantons. Such a solution may be built from a $\mathbb{Z}_2$-quotient of a configuration of calorons, which are instantons on $\mathbb{R}^3\times S^1$ that are made from pairs of BPS monopoles \cite{Lee:1998vu,Kraan:1998pm}. With the correct choice of parameters, the caloron is symmetric between pairs of points, and in the point-like limit it may provide a candidate building block for the singular gauge configuration required for this heterotic dual model. 

\subsection{Simplest $N=1$ Example }

We continue to our first $N=1$ example, which is similar in most
regards to the $N=2$ example. In this case, we have a $G_{2}$ orbifold
$X_{2}$ with $12$ $T^{3}$ of $A_{1}$ singularities and three possible
choices of K3 fibration. Although the base 3-orbifold of the fibration
differs for the different choices, our analysis of the heterotic gauge
bundle is unaffected by this change. For our analysis, we will choose
the $\alpha$-fibration, which gives the heterotic geometry $Y_{2,\alpha}=T_{123567}^{6}/\left\langle \beta,\gamma_{2}\right\rangle $
described in example 3.2.

For the perturbative part of the spectrum, in addition to the $T^{3}$
Wilson lines described above, we must choose an action of $H_{2,\alpha}=\left\langle \beta,\gamma_{2}\right\rangle $
on the perturbative gauge bundle. We choose $\beta$ to act as the
outer automorphism of $E_{8}\times E_{8}$ as in example 6.1, while
$\gamma_{2}$ must act in a way that swaps two $\text{SU}(2)^{4}$
factors within the $\text{SU}(2)^{8}$ subgroup of $E_{8}$ that is
preserved by the Wilson lines. These group actions accomplish the
monodromy seen on the $G_{2}$ orbifold side, where $\beta$ and $\gamma_{2}$
each act on the $16$ fixed loci of $\alpha$ so as to identify them
in fours. There are two $\mathbb{Z}_{2}$ elements of $E_{8}$ (corresponding
to nodes on the Dynkin diagram with Dynkin label $2$), familiar from
$T^{4}/\mathbb{Z}_{2}$ orbifolds, that are candidates for the action
of $\gamma_{2}$. The computation of the perturbative spectrum must
additionally take into account shifts in left-moving energy from point-like
instantons, as described in section 5.3. 

Now we investigate the non-perturbative spectrum. The heterotic geometry
$Y_{2,\alpha}$ has $16$ $T^{2}$ of $A_{1}$ loci, half from $\beta$
and half from $\gamma_{2}$. As in the previous example, we must produce
$\text{SU}(2)^{8}$ non-perturbative gauge symmetry from these $16$
loci. Again, the monodromy action of $\alpha$ in the ambient space
interchanges the $\beta$-loci in pairs, and now they interchange
the $\gamma_{2}$-loci in pairs as well. Thus we again expect the
gauge bundle to be made of $8$ semi-localized instantons that reside
on pairs of $T^{2}$ and come with $3$ adjoint chirals each. 

The most intuitive description of this gauge bundle configuration
(and that of the previous example) is via a ``sequential orbifold'',
where the monodromy action of $\alpha$ on the $\beta$- and $\gamma_{2}$-loci
is captured by a heterotic orbifold by the full $\Gamma_{2}$ (instead
of only the subgroup $H_{2,\alpha}$ that acts nontrivially on the
geometry). To make sense of this prescription, the elements of the
orbifold group are taken to act in a certain order, where $\alpha$
acts upon the non-perturbative $H_{2,\alpha}$-orbifold: we think
of the model as $X_{2}/\Gamma_{2}=\left(X_{2}/H_{2,\alpha}\right)/\left\langle \alpha\right\rangle $.
Because $\Gamma_{2}$ is abelian, we are free to order the elements
in this way, although a fully satisfactory interpretation of this
model would consider the non-perturbative effects of all of $\Gamma_{2}$
at once. 

Because $\alpha$ acts to swap the heterotic geometry with another
fiber of $\pi_{4}:X_{2,\alpha}\to S_{4}^{1}/\left\langle \alpha\right\rangle $,
only $H_{2,\alpha}$ descends to the heterotic geometry, which we
identify with the orbifold $Y_{2,\alpha}=T_{123567}^{6}/H_{2,\alpha}$.
Nonetheless, we may think of this string background as a $\Gamma_{2}$
background where $\alpha$ acts trivially on the geometry, but has
a nontrivial action on the gauge bundle, identifying $\text{SU}(2)$
factors in pairs. The action of $\alpha$ on the gauge bundle may
be thought of as identifying components of the connection that take
values in pairs of $\mathfrak{su}(2)$ summands. These Lie algebra
summands correspond to $\text{SU}(2)$ factors of the gauge group
that arise non-perturbatively from fixed loci of $\beta$ and $\gamma_{2}$,
so for this interpretation to reproduce the intuitive picture from
the 7D geometry, we must choose a specific order for the orbifold
actions. We construct an orbifold background on $T^{6}/H_{2,\alpha}$
with a non-perturbative spectrum from standard point-like instantons,
such as those found on the tangent sheaf, and then act on the resulting
theory with a further orbifold action by $\alpha$ that identifies
components of the resulting connection. 

Given these results, we can ask how they inform our understanding
of the half-$G_{2}$ map. In the 7D case of the half-K3 limit, the
heterotic gauge symmetry may be read off from the complicated geometry
at the ends of the interval, because all singularities were isolated,
and therefore able to be moved to the complicated ends. In our half-$G_{2}$
limit, this remains true for the perturbative gauge symmetry, since
those loci are transverse to the generic fiber, but the singular fibers
that give rise to the non-perturbative gauge symmetry necessarily
stretch all the way across the interval (see Figure 3). In the example
at hand, each singular $T^{3}$ that stretches across the interval
intersects the generic fiber in two components, while it intersects
the end fiber in only one component. This means that looking only
at the complicated ends of the interval will not determine the heterotic
gauge bundle configuration, because this information would not tell
you which pairs of $T^{2}$ loci in the heterotic geometry join into
one in the complicated end. In other words, to reconstruct the $\alpha$-monodromy,
one must look at the entire interval to follow the loci through the
6D fibers. So we conclude that the information of the heterotic gauge
bundle may be spread throughout the half-$G_{2}$ interval, even when
the metric in the bulk of the interval is trivial. 

We may again consider non-singular instanton configurations that reproduce the correct perturbative spectrum. In this case, we add a third summand to the instanton configuration:
\[
{\cal A}_{3}=\left[\left(1,1\right)\left(1,g_{R}\right)\left(1,1\right)\left(1,g_{R}\right)\left(1,1\right)\left(g_{L},1\right)\left(1,1\right)\left(g_{L},1\right);W_{3}\right]\ .
\]
Then the centralizer of the sum of ${\cal A}_{1},{\cal A}_{2},$ and ${\cal A}_{3}$ is
$\text{SU}(2)^{4}$, embedded in $\text{SO}(4)^{8}$ as 
\[
\left(1,1\right)\left(1,1\right)\left(1,1\right)\left(1,1\right)\left(1,g_{R}\right)\left(1,g_{R}\right)\left(1,g_{R}\right)\left(1,g_{R}\right)\ .
\]
Again, we get three chiral multiplets per unbroken $\text{SU}(2)$ from freedom to specify flat connections on the covering $T^6$. 

\subsection{Orbifold Singular Locus Example }

Lastly, we will look at our $N=1$ example with $T^{3}/\mathbb{Z}_{2}$
singular loci, which exhibits different point-like instanton behavior
than the previous examples and also varying bundle configurations
for different choices of fibration. We will first consider the $\alpha$-fibration,
in which case we have $8$ singular $T^{2}$ loci from $\beta$ and
an additional $8$ from $\gamma_{3}$. The $\beta$-loci come from
the intersection of $4$ $T^{3}$ loci with the heterotic geometry,
while the $\gamma_{3}$-loci come from the intersection with $8$
$T^{3}/\mathbb{Z}_{2}$ loci. So we expect $\text{SU}(2)^{4}$ gauge
symmetry with $3$ adjoint chirals per $\text{SU}(2)$ from the $8$
$\beta$-loci while we expect $\text{SU}(2)^{8}$ gauge symmetry with
only $1$ adjoint chiral per $\text{SU}(2)$ from the $8$ $\gamma_{3}$-loci.
Thus it is clear that the two loci support different types of point-like
instantons. 

We can understand the difference between the loci based on the monodromy
actions in the ambient space. The action of $\alpha$ on the $\beta$-loci
is identical to the previous example, but it does not interchange
the $\gamma_{3}$-loci, as it did for the $\gamma_{2}$-loci in the
that case. To see this, we will consider an example locus in the covering
space. The throat coordinate is $x_{4}$, and the heterotic geometry
is $Y_{3,\alpha}=\pi_{4}^{-1}\left(\frac{1}{4}\right)$. Consider
the $\gamma_{3}$-locus 
\[
L=\left(\frac{1}{4},x_{2},0,\frac{1}{4},0,x_{6},0\right) \ ,
\]
where $x_{2}$ and $x_{6}$ can vary. We must keep in mind that this
$T^{2}$ in the covering space represents the same $T^{2}$ as if
we act upon this with $\beta$: 
\[
\beta L=\left(\frac{3}{4},\frac{1}{2}-x_{2},0,\frac{1}{4},0,-x_{6},0\right) \ .
\]
Because $x_{2}$ and $x_{6}$ are free coordinates, the only change
is in the $x_{1}$ coordinate. On the other hand, we may consider
the effect of $\alpha$-monodromy on $L$. We shift along the throat
coordinate to $x_{4}=\frac{3}{4}$ and apply $\alpha$, which gives
us
\[
\alpha L_{x_{4}+\frac{1}{2}}=\left(\frac{3}{4},-x_{2},0,\frac{1}{4},0,x_{6},0\right) \ .
\]
We see that the $\alpha$-monodromy accomplishes the \emph{same} interchange
of the $\gamma_{3}$-loci in the covering space as does $\beta$,
so the action on the $\gamma_{3}$-loci in $Y_{3,\alpha}$ is trivial.
Because of this, each $T^{3}/\mathbb{Z}_{2}$ intersects the heterotic
geometry only once, and therefore the associated instantons are fully
localized on a single $T^{2}$. 

However, the monodromy of $\alpha$ does eliminate harmonic one-forms
on $T^{3}/\mathbb{Z}_{2}$ (as can be seen by the action of $\alpha\beta$
on either of the end-fibers of the $x_{4}$-interval), so that the
instanton should come with only one adjoint chiral multiplet. In $N=2$
language, the resulting gauge theory should be pure $N=2$ $\text{SU}(2)$
SYM. The existing 6D point-like instanton classification does not
appear to include a $c_{2}=3/2$ instanton that gives non-perturbative
gauge symmetry with no charged matter, so this gauge bundle configuration
may also be previously undescribed. Note that the charged matter could
be blocked by a B-field holonomy, as in \cite{Aspinwall:1998he},
but this would block the gauge symmetry as well. 

The $\beta$-fibration of $X_{3}$ gives identical results, but the
$\gamma_{3}$-fibration provides a heterotic dual with a different
gauge background. In this case, the geometry is $Y_{3,\gamma_{3}}=T_{123456}^{6}/\left\langle \alpha,\beta\right\rangle $,
which has singular loci as in example 6.2. The non-perturbative part
of the spectrum should be described, as in that case, by $8$ semi-localized
instantons on pairs of loci. The difference this time is in the perturbative
part of the compactification: as discussed for the $\alpha$-fibration,
the monodromy actions of $\alpha$ and $\beta$ on the $\gamma_{3}$
loci in the $T^{7}$ covering space are identical. Therefore, in the
$\gamma_{3}$-fibration, where the $\gamma_{3}$ loci give rise to
perturbative gauge symmetry on the heterotic side, the actions of
$\alpha$ and $\beta$ on the perturbative gauge bundle must be chosen
accordingly. In particular, if we choose $\alpha$ to act on the perturbative
gauge bundle as the outer automorphism of $E_{8}\times E_{8}$, we
must choose $\beta$ as an element of $E_{8}$ that commutes with
the resulting $\text{SU}(2)^{8}$, but reduces the charged matter
spectrum from $3$ adjoint chirals per $\text{SU}(2)$ to $1$ adjoint
chiral per $\text{SU}(2)$. 

\section{An Alternate Duality Chain via Type I}

\begin{table}
\begin{centering}
\makebox[\textwidth][c]{%
\begin{tabular}{|c|c|c|}
\hline 
 & Perturbative $\text{SU}(2)^{8}$ & Non-perturbative $\text{SU}(2)^{8}$\tabularnewline
\hline 
\hline 
M & 8 $T^{3}$ $\alpha$-loci & 8 $T^{3}$ $\beta$-loci\tabularnewline
\hline 
IIA & D6-branes on orientifold planes & 8 $T^{2}$ $\beta^{*}$-loci\tabularnewline
\hline 
I & Subgroup of D9-brane $\text{Spin}(32)/\mathbb{Z}_{2}$ & D5-branes on 16 singularities with winding shift\tabularnewline
\hline 
$\text{SO}(32)$ & Subgroup of primordial $\text{Spin}(32)/\mathbb{Z}_{2}$ & Point-like instantons on 16 singularities with winding shift\tabularnewline
\hline 
$E_{8}$ & Subgroup of primordial $E_{8}\times E_{8}$ & T-dual point-like instantons on 16 singularities\tabularnewline
\hline 
\end{tabular}}
\par\end{centering}
\caption{Origin of non-abelian gauge symmetry in the $N=2$ model at each stage
of the duality chain. ``Perturbative'' and ``Non-perturbative''
labels refer to the string coupling of the heterotic theories. }
\end{table}

To understand the gauge symmetry and particle spectrum seen in our
M-theory orbifold backgrounds, it is informative to look at another
chain of dualities that relates M-theory to the $\text{Spin}(32)/\mathbb{Z}_{2}$
heterotic string. The point-like instanton effects we have seen in
heterotic dual models look odd from the $E_{8}\times E_{8}$ perspective,
but may be better understood as $\text{Spin}(32)/\mathbb{Z}_{2}$
point-like instantons, which naturally appear with symplectic gauge
groups and without tensor multiplets. The appearance of T-dual $\text{Spin}(32)/\mathbb{Z}_{2}$
point-like instantons in $E_{8}\times E_{8}$ heterotic string theories
was found in a similar setup in \cite{Berkooz:1996iz}, where they
resolve confusions that arose from mistakenly attributing their effects
to $E_{8}\times E_{8}$ point-like instantons. They were also found
to explain the spectrum of an $E_{8}\times E_{8}$ compactification
in \cite{Aldazabal:1997wi}. Our duality chain begins with M-theory,
proceeds to a IIA orientifold, then a T-dual Type I theory, and finally
an S-dual $\text{Spin}(32)/\mathbb{Z}_{2}$ heterotic model. The latter
theory may be related to the $E_{8}\times E_{8}$ heterotic string
theory by an additional T-duality. 

\subsection{$N=2$ Example}

Beginning with our $N=2$ example of section 3.1, if we take the $x_{4}$-direction
as the M-theory circle, we may obtain a dual theory from Type IIA
on $T_{123567}^{6}$ orientifolded by the group 
\[
\Gamma_{1}^{*}=\left\langle \left(-1\right)^{F_{L}}\alpha^{*}\Omega,\beta^{*}\right\rangle =\left\langle \left(-1\right)^{F_{L}}R_{123}\Omega,R_{1234}\sigma_{2}\right\rangle \ ,
\]
where $F_{L}$ is the left-moving fermion number, $\Omega$ is the
worldsheet parity operator, $\alpha^{*}=\alpha\bigm|_{123567}$, and
similarly for $\beta^{*}$ \cite{Kachru:2001je}. We also write the
action in terms of the reflection operator $R$, which flips the coordinates
shown in its subscripts, as well as the shift operator $\sigma_{i}$ that
performs an order-two shift on coordinate $x_{i}$. In this IIA background,
an $\text{SU}(2)^{8}$ gauge symmetry arises from the D6-branes required
to cancel the RR charges created by O6-planes along the 123-directions.
An additional $\text{SU}(2)^{8}$ gauge symmetry comes from D2-branes
wrapped on the loci of $A_{1}$ singularities created by $\beta^{*}$,
which are exchanged in pairs by $\alpha^{*}$. In choosing the $x_{4}$
direction as the M-theory circle, requiring a weakly-coupled Type
IIA dual would violate the limits in which we previous formulated
our M/heterotic duality. Before, we chose the $x_{4}$ direction as
the throat direction of the half-$G_{2}$ limit and required it to
be large compared to the other dimensions of the K3 fiber. Thus, if
we want to compare our IIA model directly to M-theory in the half-$G_{2}$
limit, we must work with strong IIA coupling. We could instead choose
the $x_{7}$ direction as the M-theory circle, but this radius would
also be required to be large due to the adiabatic limit. 

Next, we apply T-duality along the 123-directions to obtain a Type
I dual. This perspective gives a conceptual advantage because the
entire spectrum is expected to be visible perturbatively on the Type
I side, and the tadpole cancellation conditions give a powerful tool
for computations. Early examples of spectrum computations using this
method include \cite{Gimon:1996rq,Gimon:1996ay,Berkooz:1996dw,Dabholkar:1996zi,Aldazabal:1998mr}.
In our case, T-duality gives Type IIB on $T_{\hat{1}\hat{2}\hat{3}567}^{6}$
orientifolded by the dual group 
\[
\tilde{\Gamma}_{1}^{*}=\left\langle \Omega,\tilde{\beta}^{*}\right\rangle =\left\langle \Omega,R_{1234}\tilde{\sigma}_{2}\right\rangle \ , 
\]
where $\tilde{\beta^{*}}$ has a winding shift in the $x_{2}$ direction
instead of the momentum shift in $\beta^{*}$ (signified by the tilde
on $\tilde{\sigma}_{2}$). The hat notation on the torus coordinates
signifies that the radii of the first three coordinates of the torus
are inverted by T-duality. The operation also transforms the D6-branes
to D9-branes that generate an $\text{SU}(2)^{8}$ gauge symmetry as
a subgroup of $\text{Spin}(32)/\mathbb{Z}_{2}$. Meanwhile, the possible
presence of D-branes at the $A_{1}$ singularities (and the resulting
gauge symmetry) is complicated by the presence of the winding shift. 

Momentum and winding shifts were originally discussed in the heterotic
context in \cite{Narain:1986qm}, and their effects were studied in
the Type I context in \cite{Antoniadis:1998ki,Antoniadis:1998ep},
where they give rise to supersymmetry breaking via stringy variants
of the Scherk-Schwarz mechanism \cite{Scherk:1978ta}. In these Type
I models, the shifts take place in directions along which the reflections
do not act. In our case, the shifts are in directions that are acted
upon by the reflection, but they cannot be removed by coordinate redefinitions.
The role of the Type I winding shift may be understood via its dual
action in the Type IIA model. Relative to the IIA model without a
shift, the momentum shift on $x_{2}$ blocks the appearance of a second
sector of D6-branes that would intersect the first sector of D6-branes.
Thus, it cuts in half the gauge symmetry and reduces the matter spectrum.
This is exactly the behavior that we want to attribute to the semi-localized
point-like instantons in the $E_{8}\times E_{8}$ heterotic dual.
Aside from the winding shift, our Type I model is similar to the
$\mathbb{Z}_{2}$-orbifold of Type I considered in \cite{Bianchi:1990tb,Gimon:1996rq}.
A variant of this model with a momentum shift was considered in \cite{Dabholkar:1996zi}. 

The last step of the duality chain is an S-duality to the $\text{Spin}(32)/\mathbb{Z}_{2}$
heterotic string. The Type I D9-brane gauge symmetry becomes the perturbative
gauge symmetry $\text{SU}(2)^{8}$ within the primordial $\text{Spin}(32)/\mathbb{Z}_{2}$
gauge group. The other $\text{SU}(2)^{8}$ is non-perturbative and
is expected to come from $\text{Spin}(32)/\mathbb{Z}_{2}$ point-like
instantons effects. The background orbifold is unchanged when passing
from Type I to the heterotic string, so the heterotic dual inherits
the winding shift, which interacts with the point-like instantons
to create the $\text{SU}(2)^{8}$ gauge symmetry. 

The $E_{8}\times E_{8}$ heterotic string may be reached by a final
T-duality between the two heterotic string theories. From this perspective,
the instanton configuration appears to be spread across two disconnected
singular loci. This duality chain provides a sequence that transforms
the geometric data from the $G_{2}$ space into the bundle data of
the $E_{8}\times E_{8}$ heterotic compactification. At the initial
M-theory stage, there are 8 singular loci that give rise to a rank-8
gauge group. In the final $E_{8}\times E_{8}$ heterotic stage, the
same rank-8 gauge group comes from 16 singular loci. In the intervening
Type I and $\text{Spin}(32)/\mathbb{Z}_{2}$ heterotic stages, the
compactification is weakly non-geometric due to the winding shift,
so there isn't a clear answer to the number of singular loci, but
the winding shift accomplishes the same rank-8 gauge group as the
initial and final stages. 

An alternative duality chain may be obtained in this $N=2$ case by
starting with a different Type IIA limit. Our M-theory background
is $T^{7}/\left\langle \alpha,\beta\right\rangle $, where none of
the elements in the orbifold group act on the final coordinate, $x_{7}$.
Thus, we may take this coordinate as the M-theory circle and obtain
a IIA dual on $T_{123456}^{6}/\left\langle \alpha,\beta\right\rangle $,
which is again the orbifold limit of the Borcea--Voisin manifold of
Hodge numbers $(19,19)$. The geometric limits discussed in section
3 require that the radius of $x_{7}$ is large, meaning that this
IIA dual is strongly-coupled. For our purposes, the only relevant
non-perturbative effects are the massless states that arise from wrapped
D2-branes on the orbifold singularities. 

Type I and heterotic duals to this model were considered in \cite{Gregori:2001ak},
where it was found that the Type I dual includes momentum or winding
shifts along the invariant $T^{2}$. This is in contrast to the Type
I duals found in our duality chain above, where these shifts were
along a direction of a $T^{4}$ on which the orbifold group acts nontrivially.
The massless states in the heterotic dual of \cite{Gregori:2001ak}
were found to all be of non-perturbative origin, suggesting that this
heterotic dual is distinct from the one obtained in the half-$G_{2}$
limit, which has a mixture of perturbative and non-perturbative gauge
symmetry. This second duality chain is not available in the $N=1$
cases, because there is no coordinate on which the M-theory orbifold
group acts trivially, so we may not obtain a IIA orbifold dual in
the same manner. 

An additional Type IIB dual may be obtained by applying T-duality
along only the $x_{3}$-direction instead of the $x_{123}$-directions.
In this case, we find Type IIB compactified on $T_{12\hat{3}567}^{6}/\left\langle \Omega R_{12},R_{1234}\sigma_{2}\right\rangle $.
Cancellation of the O7-plane charge created at fixed points of $\Omega R_{12}$
will create a D7-brane background, so this dual model should be expressible
in terms of F-theory, along the lines of \cite{Braun:2017uku}. 

\subsection{The $N=1$ Examples}

In the $N=1$ cases, we also must take into account the nontrivial
action of $\gamma$ as we go through the steps of the duality chain.
A similar Type I orbifold was studied in \cite{Berkooz:1996dw}, and
further examples are given in \cite{Aldazabal:1997wi,Aldazabal:1998mr}.
A similar duality chain was considered for M-theory on $\text{Spin}(7)$
orbifolds in \cite{Majumder:2001dx}. Our model differs from that
of \cite{Berkooz:1996dw} by the inclusion of winding shifts in multiple
directions that avoid an intersecting brane interpretation and reduce
the rank of the gauge symmetry. In the $N=1$ cases, discrete torsion
is a nontrivial choice in the orbifold backgrounds as well. In our
cases, it is expected to be present, as in \cite{Acharya:1996fx}. 

For the IIA dual of our M-theory model on $T^{7}/\left\langle \alpha,\beta,\gamma_{2}\right\rangle $
of example 3.2, we take $x_{4}$ to be the M-theory direction, so
that we obtain the dual theory IIA on $T_{123567}^{6}$ orientifolded
by
\[
\Gamma_{2}^{*}=\left\langle \left(-1\right)^{F_{L}}\alpha^{*}\Omega,\beta^{*},\gamma_{2}^{*}\right\rangle =\left\langle \left(-1\right)^{F_{L}}R_{123}\Omega,R_{1256}\sigma_{2},R_{1357}\sigma_{1}\sigma_{3}\right\rangle \ .
\]
This is the dual model labeled as ``Orientifold B'' in \cite{Kachru:2001je}.
Applying T-duality in the 123-directions gives us Type IIB on $T_{\hat{1}\hat{2}\hat{3}567}^{6}$
orientifolded by 

\[
\tilde{\Gamma}_{2}^{*}=\left\langle \Omega,\tilde{\beta}^{*},\tilde{\gamma}_{2}^{*}\right\rangle =\left\langle \Omega,R_{1256}\tilde{\sigma}_{2},R_{1357}\tilde{\sigma}_{1}\tilde{\sigma}_{3}\right\rangle \ .
\]
The winding shifts persist in the S-dual $\text{Spin}(32)/\mathbb{Z}_{2}$
heterotic model as well. If we apply T-duality to convert this to
an $E_{8}\times E_{8}$ heterotic model, we end up with an instanton
configuration that looks locally similar to the $N=2$ case. 

The M-theory background of example 3.3, which lives on the space $T^{7}/\left\langle \alpha,\beta,\gamma_{3}\right\rangle $,
is similarly dual to Type IIB on $T_{123567}^{6}$ orientifolded by 

\[
\tilde{\Gamma}_{3}^{*}=\left\langle \Omega,\tilde{\beta}^{*},\tilde{\gamma}_{3}^{*}\right\rangle =\left\langle \Omega,R_{1256}\tilde{\sigma}_{2},R_{1357}\tilde{\sigma}_{1}\right\rangle \ ,
\]
where the only difference from the previous example is the lack of
a winding in the $x_{3}$-direction. Thus, while the instantons in
models 6.2 and 6.3 look rather different from the $E_{8}\times E_{8}$
heterotic perspective, the models differ on the $\text{Spin}(32)/\mathbb{Z}_{2}$
side only by the inclusion of a winding shift on one coordinate, just
as they differed on the M-theory side by only a momentum shift. Explicit
calculations of the effect of winding shifts on the $T^{6}/\mathbb{Z}_{2}^{2}$
background of \cite{Berkooz:1996dw} would further explain the instanton
effects, but is beyond the scope of this work. 

\section{Discussion}

To better understand the types of point-like instantons that appear
in our $E_{8}\times E_{8}$ backgrounds, we may compare examples 6.2
and 6.3, our two $N=1$ cases. These examples live on the same Calabi--Yau orbifold, so the difference in their non-perturbative gauge symmetry
cannot come from any mechanism that depends on the geometry alone.
For example, one might expect that the superpotential contributions
from worldsheet instantons could lift gauge bundle moduli in a way
that differentiates the two cases. However, the presence of worldsheet
instanton effects at lowest order is controlled only by the existence
of rigid rational curves, so it is a property only of the geometry
\cite{Harvey:1999as}. Thus, if we are to appeal to some part of the
heterotic background to explain the differences in non-perturbative
behavior, it must be the background gauge field or B-field. A particularly
attractive mechanism is Wilson line backgrounds. We have already specified
the perturbative Wilson line background via the half-K3 limit, but
there may be additional Wilson line effects involving the non-perturbative
part of the gauge group, and these may break this part of the gauge
symmetry in the low energy effective theory. To further understand
the behavior of the non-perturbative spectra in our examples, we will
discuss the relation to two other heterotic phenomena: Ho\v{r}ava--Witten
duals and coupled heterotic moduli.

\subsection{Gauge Locking in Ho\v{r}ava--Witten Duals}

As observed in \cite{Acharya:1996ci}, Ho\v{r}ava--Witten theory \cite{Horava:1995qa,Horava:1996ma}
suggests that our heterotic models should have an additional M-theory
dual on a background of the form $T^{6}/H\times S^{1}/\mathbb{Z}_{2}$.
Then, via the heterotic string, we should have an M-theory/M-theory
duality between compactifications on $G_{2}$ spaces and Ho\v{r}ava--Witten
compactifications. One interesting aspect of this duality is how the
heterotic point-like instantons are represented on each side. In the
heterotic duality with Ho\v{r}ava--Witten theory, point-like instantons
on orbifold singularities are thought of as fractional M5-branes wrapped
on the singularity. On the other hand, in the duality with M-theory
on $G_{2}$, the instantons correspond to M2-branes wrapped on degenerate
K3 fibers. This is an example of electromagnetic duality for the C-field
that interchanges M2 and M5 branes \cite{Duff:1996rs,Acharya:1996ea}.
Thus, Ho\v{r}ava--Witten theory offers an electromagnetically dual perspective
from which to investigate our phenomena. 

In the dual pairs of examples 6.1 and 6.2, we found that the M-theory
geometry dictates a spectrum that looks subtle from the $E_{8}\times E_{8}$
heterotic side, where gauge symmetries from different singular loci
are united. This phenomenon is familiar from studies of heterotic
orbifolds via Ho\v{r}ava--Witten theory, where it has been found that 7-planes
stretching between the 10-plane ends of the M-theory interval can
carry gauge degrees of freedom that ``lock'' together, reducing
to a smaller subgroup \cite{Kaplunovsky:1999ia,Gorbatov:2001pw,Marquart:2002bz,Claussen:2016ucd}.
An example considered first in \cite{Kaplunovsky:1999ia} and later
in \cite{Marquart:2002bz} is a heterotic compactification on $T^{4}/\mathbb{Z}_{2}$
with perturbative gauge group $\text{SO}(16)\times E_{7}\times\text{SU}(2)$
(up to $\mathbb{Z}_{2}$ quotients). The point-like instantons required
to cancel the magnetic charge of the 16 $A_{1}$ singularities would
naively contribute a non-perturbative gauge symmetry of $\text{SU}(2)^{16}$,
but it can be shown by duality with F-theory that all $\text{SU}(2)$
factors are broken to a common diagonal $\text{SU}(2)$, denoted $\text{SU}(2)^{*}$,
so that the full gauge group is $\text{SO}(16)\times E_{7}\times\text{SU}(2)^{*}$.
In this sense, all of the non-perturbative $\text{SU}(2)$ factors
and the perturbative $\text{SU}(2)$ factor are ``locked'' together.
The M-theory mechanism invoked to describe this phenomenon is nonzero
$G$-flux required by anomaly cancellation, deforming the Ho\v{r}ava--Witten
geometry away from a metric product. The gauge locking explains how
the perturbative twisted spectrum can include matter charged under
both $E_{8}$ factors, even though they are separated at either end
of the Ho\v{r}ava--Witten interval: the singular 7-planes carry the
gauge quantum numbers between the two ends. 

In \cite{Claussen:2016ucd}, similar phenomena were found for the
Ho\v{r}ava--Witten picture of a heterotic $T^{6}/\mathbb{Z}_{3}$ orbifold.
In this case, the effective theory is 4D and the states charged under
the two $E_{8}$ factors are not localized to one side. Instead, the
states that carry the bifundamental representation of $\text{SU}(3)$
subgroups of the two $E_{8}$ factors are spread over the length of
the interval in a meson-like configuration. 

These Ho\v{r}ava--Witten phenomena---gauge locking and delocalized bundle
configurations---are very similar to the semi-localized instantons
that we observe in our examples, so it is possible that they are incarnations
of the same type of phenomenon seen from dual perspectives.
However, our examples do not have a topological defect analogous to
an orbifold 7-plane to carry quantum numbers between matter loci.
Additionally, the gauge locking is achievable on heterotic backgrounds
that lack a momentum shift, so its interpretation in a dual Type I
model may be quite different from that of the semi-localized instantons.
The relation between these phenomena is an interesting question for
future work. 

\subsection{Coupled Heterotic Moduli}

An important feature of heterotic compactifications is that the moduli
space does not factorize into complex structure and gauge bundle moduli:
the two are coupled by the fact that the gauge bundle must remain
holomorphic, so that a particular bundle configuration is compatible
with only certain deformations of the complex structure \cite{Anderson:2011ty}.
This may allow our semi-localized point-like instantons to lift moduli
that are unphysical from the M-theory perspective by coupling bundle
moduli to the K{\"a}hler and complex structure moduli of the loci on which
they are supported. For instance, in example 6.2, because the $T^{3}$
loci of the $G_{2}$ orbifold intersect the heterotic geometry in
$T^{2}\sqcup T^{2}$ loci, the $T^{2}$ loci cannot be blown up or
deformed independently but must have their moduli coupled, as they
are part of the same $T^{3}$ locus in the ambient space. Thus, coupling
of these moduli by semi-localized instantons of the gauge bundle looks
quite natural. In this sense, we may think of the singularities of
the heterotic orbifold as ``partially frozen'', since the directions
of moduli space that correspond to independent resolutions of singular
loci have become massive. 

\subsection{Future Directions}

This paper is based on the half-$G_{2}$ limit and point-like instantons
on orbifold singularities, neither of which has been fully understood
in the literature. Consequently, there are many directions in which
this work can be taken to deepen our knowledge of non-perturbative
aspects of M/heterotic duality. 
\begin{itemize}
\item As discussed in previous sections, there are several perturbative
and non-perturbative spectrum computations that would elucidate the
relations between our M-theory, heterotic, and Type I backgrounds,
but were beyond the scope of this work. Of particular interest would
be a calculation of the Type I spectra with the effects of winding
shifts, as described in section 7, as well as a calculation of the
heterotic spectra taking into account Wilson lines and the lack of
modular invariance, as in \cite{Aldazabal:1997wi}. 
\item In this paper, we restricted ourselves to $A_{1}$ singularities,
but there exist examples of $G_{2}$ orbifolds with other ADE singularities.
How does the half-$G_{2}$ map operate in those situations? The choice
of a throat coordinate was made simple by the fact that the elements
of $\Gamma$ acted as reflections, but the choice may not be so obvious
if the group elements act in more complicated ways. 
\item A next step in the understanding of the half-$G_{2}$ map would be
to consider more general M-theory backgrounds that include nontrivial
profiles for the C-field and 7D gauge fields. Additionally, studying
$G_{2}$ orbifolds with intersecting codimension $4$ singularities
and/or codimension $7$ singularities will allow for a greater variety
of matter representations. The Type I tadpole cancellation conditions
in the alternate duality chain of section 7 give another way to look
at the presence or absence of singularities in the $G_{2}$ moduli
space. 
\item The examples of $G_{2}$ orbifolds that we look at in this paper are
non-generic in the sense that they have \emph{multiple }K3 fibrations,
giving us extra tools to work with in determining the heterotic gauge
bundle. In particular, extra K3 fibrations on the M-theory side will
guarantee a K3-fibration on the heterotic side (in the half-$G_{2}$
limit), which simplifies our treatment of point-like instantons by
increasing the amount of supersymmetry in the local theory. Eventually,
the half-$G_{2}$ map should be generalized to K3-fibered $G_{2}$
orbifolds that have only one fibration and dual heterotic orbifolds
that only enjoy an SYZ fibration.
\item Reconstruction of heterotic gauge bundles from fiberwise data on a
$T^{3}$ fibration is not yet well-understood, but progress is being
made in that direction via the 3D Hitchin system and related spectral
cover descriptions of heterotic gauge bundles \cite{Pantev:2009de,Braun:2018vhk,Barbosa:2019bgh,Acharya:2020xgn}.
These methods give a promising route toward a rigorous algorithm for
constructing non-perturbative heterotic duals of M-theory backgrounds. 
\end{itemize}

\section*{Acknowledgements}

The authors would like to thank 
Paul Aspinwall,
Rodrigo Barbosa,
Andreas Braun, 
Seth Koren,
and
Eirik Eik Svanes 
for helpful conversations during the course
of this work. The work of BSA is supported by grant \#488569 (Acharya) from the Simons Foundation. AK and DRM are supported by the Simons Foundation Grant \#488629 (Morrison).
\bibliographystyle{amsunsrt-enspa}
\bibliography{G2_references_v2}

\end{document}